\documentclass[12pt,preprint]{aastex}
\usepackage{graphicx}
\usepackage{lscape}
\usepackage{enumerate}
\usepackage{xcolor}

\newcommand{\bdv}[1]{\mbox{\boldmath$#1$}}

\def\au{{\rm au}}

\def\masyr{{\rm mas}\,{\rm yr}^{-1}}
\def\kpc{{\rm kpc}}
\def\mas{{\rm mas}}

\def\muas{\mu{\rm as}}

\def\rel{{\rm rel}}

\def\e{{\rm E}}
\def\bpi{{\bdv\pi}}
\def\bmu{{\bdv\mu}}

\def\bgamma{{\bdv\gamma}}

\begin{document}
\title{Systematic KMTNet Planetary Anomaly Search. VI.  Complete Sample of 2018 Sub-Prime-Field Planets}

\author{
Youn Kil Jung$^{1,2}$, Weicheng Zang$^{3}$, Cheongho Han$^{4}$, Andrew Gould$^{5,6}$, Andrzej Udalski$^{7}$,
(Lead Authors)\\
Michael D. Albrow$^{8}$, Sun-Ju Chung$^{1}$, Kyu-Ha Hwang$^{1}$, Yoon-Hyun Ryu$^{1}$, In-Gu Shin$^{4}$, 
Yossi Shvartzvald$^{9}$, Hongjing Yang$^{3}$, Jennifer C. Yee$^{10}$, Sang-Mok Cha$^{1,11}$, Dong-Jin Kim$^{1}$,
Seung-Lee Kim$^{1}$, Chung-Uk Lee$^{1}$, Dong-Joo Lee$^{1}$, Yongseok Lee$^{1,11}$, Byeong-Gon Park$^{1,2}$, 
Richard W. Pogge$^{6}$ \\
(The KMTNet Collaboration)\\
Przemek Mr{\'o}z$^{7}$, Micha{\l} K. Szyma{\'n}ski$^{7}$, Jan Skowron$^{7}$, Radek Poleski$^{7}$, 
Igor Soszy{\'n}ski$^{7}$, Pawe{\l} Pietrukowicz$^{7}$, Szymon Koz{\l}owski$^{7}$, Krzysztof Ulaczyk$^{12}$, 
Krzysztof A. Rybicki$^{7,9}$, Patryk Iwanek$^{7}$, Marcin Wrona$^{7}$\\
(The OGLE Collaboration)\\
}

\affil{$^{1}$Korea Astronomy and Space Science Institute, Daejon
34055, Republic of Korea}

\affil{$^{2}$Korea University of Science and Technology, Korea, 
(UST), 217 Gajeong-ro, Yuseong-gu, Daejeon, 34113, Republic of Korea}

\affil{$^{3}$ Department of Astronomy,
Tsinghua University, Beijing 100084, China}

\affil{$^{4}$Department of Physics, Chungbuk National University,
Cheongju 28644, Republic of Korea}

\affil{$^{5}$Max-Planck-Institute for Astronomy, K\"{o}nigstuhl 17,
69117 Heidelberg, Germany}

\affil{$^{6}$Department of Astronomy, Ohio State University, 140 W.
18th Ave., Columbus, OH 43210, USA}

\affil{$^{7}$Astronomical Observatory, University of Warsaw, 
Al.~Ujazdowskie~4, 00-478~Warszawa, Poland}

\affil{$^{8}$University of Canterbury, Department of Physics and
Astronomy, Private Bag 4800, Christchurch 8020, New Zealand}

\affil{$^{9}$Department of Particle Physics and Astrophysics, 
Weizmann Institute of Science, Rehovot 76100, Israel}

\affil{$^{10}$ Center for Astrophysics $|$ Harvard \& Smithsonian, 60 Garden
St., Cambridge, MA 02138, USA}

\affil{$^{11}$School of Space Research, Kyung Hee University,
Yongin, Kyeonggi 17104, Republic of Korea}

\affil{$^{12}$Department of Physics, University of Warwick, 
Gibbet Hill Road, Coventry, CV4~7AL,~UK}

\begin{abstract}

We complete the analysis of all 2018 sub-prime-field microlensing planets
identified by the KMTNet AnomalyFinder.  Among the 9 previously unpublished 
events with clear planetary solutions, 6 are clearly planetary
(KMT-2018-BLG-0030,
KMT-2018-BLG-0087,
KMT-2018-BLG-0247,
OGLE-2018-BLG-0298,
KMT-2018-BLG-2602, and
OGLE-2018-BLG-1119), while the remaining 3 are ambiguous in nature.
In addition, there are 8 previously published sub-prime field planets
that were selected by the AnomalyFinder algorithm.
Together with a companion paper \citep{2018prime} 
on 2018 prime-field planets, this work lays the basis for the
first statistical analysis of the planet mass-ratio function based on
planets identified in KMTNet data.  As expected \citep{zhu14},
half (17/33) of the 2018 planets likely to enter the mass-ratio analysis
have non-caustic-crossing anomalies.  However, only 1 of the 5
non-caustic anomalies with planet-host mass ratio $q<10^{-3}$ was discovered
by eye (compared to 7 of the 12 with $q>10^{-3}$), 
showing the importance of the semi-automated AnomalyFinder search.

\end{abstract}


\section{{Introduction}
\label{sec:intro}}

This paper completes the publication of all planetary events that were
identified by the KMTNet AnomalyFinder algorithm \citep{ob191053,af2}
that occurred during the 2018 season within the 21 sub-prime KMTNet fields.
It is a companion to a paper on the 2018 prime-field planets
by \citet{2018prime}, which analyzed 10 new planetary (or potentially planetary)
events, and summarized 4 previous AnomalyFinder 2018 prime-field
discoveries \citep{ob180383,kb190253}, as well as 12 previously
analyzed planetary (or possibly planetary)
events that were recovered by AnomalyFinder.  These 26
events are listed in their Table~11.  The above references are, respectively,
Papers I, IV, II, III, and V, in the AnomalyFinder series.  The locations
and cadences of the KMTNet fields are shown in Figure~12 of \citet{eventfinder}.
The prime fields are those with nominal cadences of $\Gamma=2\,{\rm hr}^{-1}$,
namely, BLG01, BLG02, BLG03, BLG41, BLG42, and BLG43.  We label the (7, 10, 3) 
remaining fields, with respective nominal cadences 
$\Gamma = (1.0, 0.4, 0.2)\,{\rm hr}^{-1}$, as
``sub-prime''. 

The AnomalyFinder \citep{af2}
identified a total of 173 anomalous events (from an underlying sample of 1728
sub-prime-field events), which it classified as
``planet'' (17), ``planet/binary'' (4), ``binary/planet'' (19),
``binary'' (126), and ``finite source'' (7).
Among the 126 in the ``binary'' classification, 35 were judged by
eye to be unambiguously non-planetary in nature.  Among the 17 in the
``planet'' classification, 7 were either previously published (5)
or in preparation (2).  Among the 4 in the ``planet/binary'' classification,
one was in preparation,
and among the 19 in the ``binary/planet'' classification 
one was a previously published planet.  None of the events that were
classified as ``binary'' or ``finite source'' were previously published
(or in preparation) planets.

The results from \citet{2018prime} 
and this paper can be combined with a detection efficiency 
analysis (Zang, Jung et al., in preparation) to derive the first
mass-ratio function based on the KMTNet project.  We refer
the reader to the introduction of \citet{2018prime} 
for the general framework of this approach.

\section{{Observations}
\label{sec:obs}}

The description of the observations is nearly identical to that in 
\citet{2018prime}
except that the events analyzed here are derived from 1728 sub-prime
events that were subjected to the AnomalyFinder algorithm compared
to 843 prime-field events in \citet{2018prime}.
In particular, the KMTNet data are taken from three identical 1.6m telescopes,
each equipped with cameras of 4 deg$^2$ \citep{kmtnet} 
and located in Australia (KMTA),
Chile (KMTC), and South Africa (KMTS).  When available, our general policy is
to include Optical Gravitational Lensing Experiment (OGLE) and 
Microlensing Observations in Astrophysics (MOA) data in the analysis.
However, none of the 9 events analyzed here were alerted by MOA.
OGLE data were taken using their 1.3m telescope
with $1.4\,{\rm deg}^2$ field of view at Las Campanas Observatory in Chile.
For the light-curve analysis, we use only the $I$-band data.

As in \citet{2018prime}, Table~\ref{tab:names} gives basic
observational information about each event.  Column~1 gives the
event names in the order of discovery (if discovered by multiple teams),
which enables cross identification.  The nominal cadences are given in column 2,
and column 3 shows the first discovery date.  The remaining four columns
show the event coordinates in the equatorial and galactic systems.
Events with OGLE names were originally discovered by the OGLE Early
Warning System \citep{ews1,ews2}.  KMT-named events with alert dates
were originally discovered by the KMT AlertFinder system \citep{alertfinder},
while the others were discovered post-season by the EventFinder system 
\citep{eventfinder}.

To the best of our knowledge, there were no ground-based follow-up observations
of any of these events.  KMT-2018-BLG-0173 was observed by {\it Spitzer}
as part of a large-scale microlensing program \citep{yee15}, but these
data do not show a discernible signal.

The KMT and OGLE data were reduced using difference image analysis 
\citep{tomaney96,alard98},
as implemented by each group, i.e., \citet{albrow09} and
\citet{wozniak2000}, respectively.  

\section{{Light Curve Analysis}
\label{sec:anal}}

\subsection{{Preamble}
\label{sec:anal-preamble}}

We present here a compressed version of Section~3.1 of \citet{2018prime}
of the common features of the light-curve analysis.  The reader who
is interested in more details should consult that work.

All of the events can be initially approximated by 1L1S models, which
are specified by three \citet{pac86}
parameters, $(t_0,u_0,t_\e)$, i.e., the time of lens-source closest
approach, the impact parameter in units of $\theta_\e$ and the Einstein
timescale,
\begin{equation}
t_\e = {\theta_\e\over\mu_\rel}; \qquad
\theta_\e = \sqrt{\kappa M\pi_\rel}; \qquad
\kappa\equiv {4\,G\over c^2\,\au} \simeq 8.14\,{\mas\over M_\odot},
\label{eqn:tedef}
\end{equation}
where $M$ is the lens mass, $\pi_\rel$ and $\bmu_\rel$ are the
lens-source relative parallax and proper-motion, respectively,
and $\mu_\rel \equiv |\bmu_\rel|$.  The notation ``$n$L$m$S'' means $n$ lenses
and $m$ sources.   In addition, to these 3 non-linear parameters, there are
2 flux parameters, $(f_S,f_B)$, that are required for each observatory,
representing the source flux and the blended flux.

We then search for ``static'' 2L1S solutions, which generally require 4
additional parameters $(s,q,\alpha,\rho)$, i.e., the planet-host separation
in units of $\theta_\e$, the planet-host mass ratio, the angle of the
source trajectory relative to the binary axis, and the angular source
size normalized to $\theta_\e$, i.e., $\rho=\theta_*/\theta_\e$.

We first conduct a grid search with $(s,q)$ held fixed at a grid of values
and the remaining 5 parameters allowed to vary in a Monte Carlo Markov chain
(MCMC). After we identify one or more local minima, we refine these by
allowing all 7 parameters to vary.

We often make use of the   heuristic analysis introduced by \citet{kb190253}
and modified by \citet{kb211391} based on further investigation in 
\citet{2018prime}.
If a brief anomaly at $t_{\rm anom}$
is treated as due to the source crossing the planet-host axis,
then one can estimate two relevant parameters
\begin{equation}
s^\dagger_\pm = {\sqrt{4 + u_{\rm anom}^2}\pm u_{\rm anom}\over 2}; \quad
\tan\alpha ={u_0\over\tau_{\rm anom}},
\label{eqn:heuristic}
\end{equation}
where $u_{\rm anom}^2= \tau_{\rm anom}^2 + u_0^2$ and
$\tau_{\rm anom} = (t_{\rm anom}-t_0)/t_\e$.
Usually,
$s^\dagger_+>1$ corresponds to anomalous bumps and
$s^\dagger_-<1$ corresponds to anomalous dips.   
This formalism predicts that if there are two degenerate solutions, $s_{\pm}$,
then they both have the same $\alpha$ and that there exists a $\Delta\ln s$
such that
\begin{equation}
s_\pm = s^\dagger_{\rm pred}\exp(\pm \Delta \ln s),
\label{eqn:heuristic2}
\end{equation}
where $\alpha$ and $s^\dagger$ are given by Equation~(\ref{eqn:heuristic}).
To test this prediction in individual cases, we can compare the purely
empirical quantity $s^\dagger\equiv \sqrt{s_+ s_-}$ with prediction
from Equation~(\ref{eqn:heuristic}), which we always label with a subscript,
i.e., either $s^\dagger_+$ or $s^\dagger_-$.  This formalism can also be used
to find ``missing solutions'' that have been missed in the grid search,
as was done, e.g., for the case of KMT-2021-BLG-1391 \citep{kb211391}.

For cases in which the anomaly is a dip, the mass ratio $q$ can be estimated,
\begin{equation}
q = \biggl({\Delta t_{\rm dip}\over 4\, t_\e}\biggr)^2
{s^\dagger\over |u_0|}|\sin^3\alpha|,
\label{eqn:qeval}
\end{equation}
where $\Delta t_{\rm dip}$ is the full duration of the dip.  
In some cases, we investigate whether the microlens parallax vector,
\begin{equation}
\bpi_\e\equiv {\pi_\rel\over \theta_\e}\,{\bmu_\rel\over\mu_\rel}
\label{eqn:piedef}
\end{equation}
can be constrained by the data.  When both $\pi_\e$ and $\theta_\e$ are
measured, they can be combined to yield,
\begin{equation}
M = {\theta_\e\over\kappa\pi_\e}; \qquad
D_L = {\au\over \theta_\e\pi_\e + \pi_S},
\label{eqn:mpirel}
\end{equation}
where $D_L$ is the distance to the lens and
$\pi_S$ is the parallax of the source.

To model the parallax effects due to Earth's orbital motion, we add
two parameters $(\pi_{\e,N},\pi_{\e,E})$, which are the components of
$\bpi_\e$ in equatorial coordinates.  We also add (at least initially)
two parameters $\bgamma =[(ds/dt)/s,d\alpha/dt]$, where
$s\bgamma$ are the first derivatives of projected lens orbital position 
at $t_0$, i.e., parallel and perpendicular to the projected separation of the
planet at that time, respectively.
In order to eliminate unphysical solutions, we impose a  constraint
on the ratio of the transverse kinetic to potential energy,
\begin{equation}
\beta \equiv \bigg|{\rm KE\over PE}\bigg|
= {\kappa M_\odot {\rm yr}^2\over 8\pi^2}\,{\pi_\e\over\theta_\e}\gamma^2
\biggr({s\over \pi_\e + \pi_S/\theta_\e}\biggr)^3 < 0.8 .
\label{eqn:betadef}
\end{equation}

It often happens that $\bgamma$ is neither significantly constrained
nor significantly correlated with $\bpi_\e$.  In these cases, we suppress
these two degrees of freedom.

Particularly if there are no sharp caustic-crossing features in the light curve,
2L1S events can be mimicked by 1L2S events.  Where relevant, we test for
such solutions by adding at least 3 parameters 
$(t_{0,2},u_{0,2},q_F)$ to the 1L1S models.
These are the time of closest approach and impact parameter of the
second source and the ratio of the second to the first source flux
in the $I$-band. 
If either lens-source approach can be interpreted
as exhibiting finite source effects, then we must add one or two further
parameters, i.e., $\rho_1$ and/or $\rho_2$.  And, if the two sources
are projected closely enough on the sky, one must also consider
source orbital motion.

In a few cases, we make kinematic arguments that solutions are unlikely
because their inferred proper motions $\mu_\rel$ are too small.   
These arguments rely on the fact that the fraction of events with
proper motions less than a given $\mu_\rel\ll \sigma_\mu$ is 
\begin{equation}
p(\leq\mu_\rel) ={(\mu_\rel/\sigma_\mu)^3\over 6\sqrt{\pi}}
\rightarrow 4\times 10^{-3}\biggl({\mu_\rel\over1\,\masyr}\biggr)^3,
\label{eqn:probmu}
\end{equation}
where (following \citealt{gould21})
we approximate the bulge proper motions as an isotropic Gaussian
with dispersion $\sigma_\mu = 2.9\,\masyr$.
For example, $p(\leq 0.5\,\masyr) = 5\times 10^{-4}$ and
$p(\leq 0.1\,\masyr) = 4\times 10^{-6}$.

\subsection{{KMT-2018-BLG-0030}
\label{sec:anal-kb180030}}

\begin{figure}
\epsscale{1.0}
\plotone{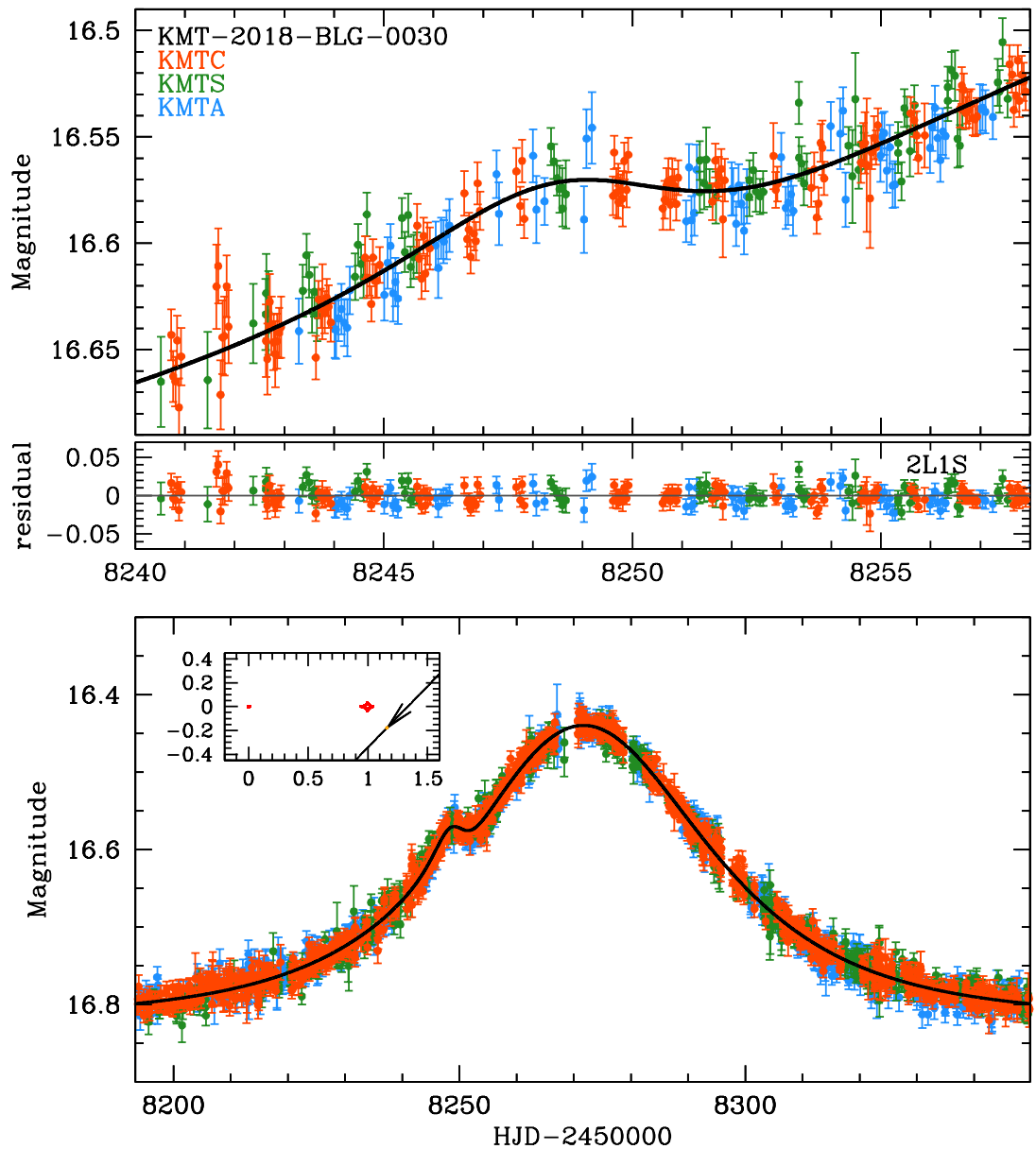}
\caption{Data (color-coded by observatory) together with the prediction
and residuals for the model of KMT-2018-BLG-0030 specified in 
Table~\ref{tab:kb0030parms}.  The short ``bump'' at $t_{\rm anom}= 8248.0$
is caused by the source crossing a ridge extending from the planetary
caustic due to a $\log q= -2.5$ super-Jovian mass-ratio planet.  See inset.
}
\label{fig:0030lc}
\end{figure}

Figure~\ref{fig:0030lc} shows a low-amplitude 
$(\Delta I\simeq ~0.4)$ microlensing
event, peaking at $t_0=8271.46$ and punctuated by a short bump at 
$t_{\rm anom}\simeq 8248.0$, i.e., $-23.46$ days before peak.  Assuming that
the source is unblended (as is reasonable, see below), the remaining
\citet{pac86} parameters are $u_0=0.90$ and $t_\e=28\,$days.  Then
$\tau_{\rm anom} = -0.84$ and $u_{\rm anom} = 1.23$.  Hence, 
Equation~(\ref{eqn:heuristic}) predicts $s^\dagger_+=1.79$ and $\alpha=133^\circ$.

We initially (as usual) conduct the grid search with free blending.  
This yields a result that is consistent with zero blending, but with a 
relatively large error, $f_B/f_S\simeq -0.14\pm 0.14$.  The extinction
from the KMT website (ultimately derived from \citealt{gonzalez12}, using
$A_I = 7\,A_K$), $A_I=3.25$, implies a dereddened baseline magnitude 
that is brighter than the clump.  The distribution of blending fractions 
for microlensing events associated with such very bright stars is generally
bimodal, i.e., either very low because the apparent bright star is the source, 
or very high because a random field star is the source.
Hence, all evidence is consistent
with very low blending and we impose zero blending.  The grid search
returns only a single solution, whose refinement is shown in 
Table~\ref{tab:kb0030parms}.
The resulting $\alpha=134.8^\circ$ is in reasonable accord the heuristic
prediction, while the value of $s=s_{\rm outer}=1.58$, would seem to suggest
a second solution at $s_{\rm inner} = (s^\dagger_+)^2/s_{\rm outer}=2.03$.
That is, Figure~\ref{fig:0030lc} shows the source passing ``outside''
the planetary caustic, so that the ``inner/outer degeneracy''
\citep{gaudi97} would seem to suggest a second solution with the source
passing inside the planetary caustic.  We specifically search for 
such a solution, in case that it was somehow missed by the grid search,
by seeding the alternate parameters suggested by the heuristic analysis.
We locate an ``inner'' solution, but it is disfavored by $\Delta\chi^2=208$,
thus confirming its rejection at the grid-search stage.

As shown in Figure~\ref{fig:0030lc}, the bump is featureless, so that
it could in principle be generated by a second source rather than a second
lens \citep{gaudi98}.  We therefore investigate 1L2S solutions 
but find that these are excluded by $\Delta\chi^2=100.5$.  

Such featureless bumps can also, in principle, be caused by a large
source passing over one of the two caustics due to a minor-image
perturbation (i.e., $s<1$).  As mentioned above, the grid search
did not return any such solution.  Nevertheless, as a matter of due diligence,
we specifically search for these.  However, the best one has extreme
negative blending $(f_B/(f_B+f_S)) = -5.5$, and $\Delta\chi^2=34$.
When we enforce zero blending, the resulting models do not even approximate
the observed light curves.  Thus, $s<1$ solutions are ruled out.

Table~\ref{tab:kb0030parms} 
indicates that this is a super-Jovian mass-ratio planet,
$\log q=-2.56$.

Due to the low amplitude of the event (and so, relatively poorly
constrained blending in a free fit), we do not attempt to measure
the microlens parallax, $\bpi_\e$.  Finally, we note that 
$\rho<0.112$ (at $2.5\,\sigma$) is only weakly constrained.
We will give quantitative expression to this weakness in
Sections~\ref{sec:cmd-kb180030} and \ref{sec:phys-kb180030}.

KMT-2018-BLG-0030 is one of three previously known planets that
are analyzed here for the first time.

\subsection{{KMT-2018-BLG-0087}
\label{sec:anal-kb180087}}

\begin{figure}
\epsscale{1.0}
\plotone{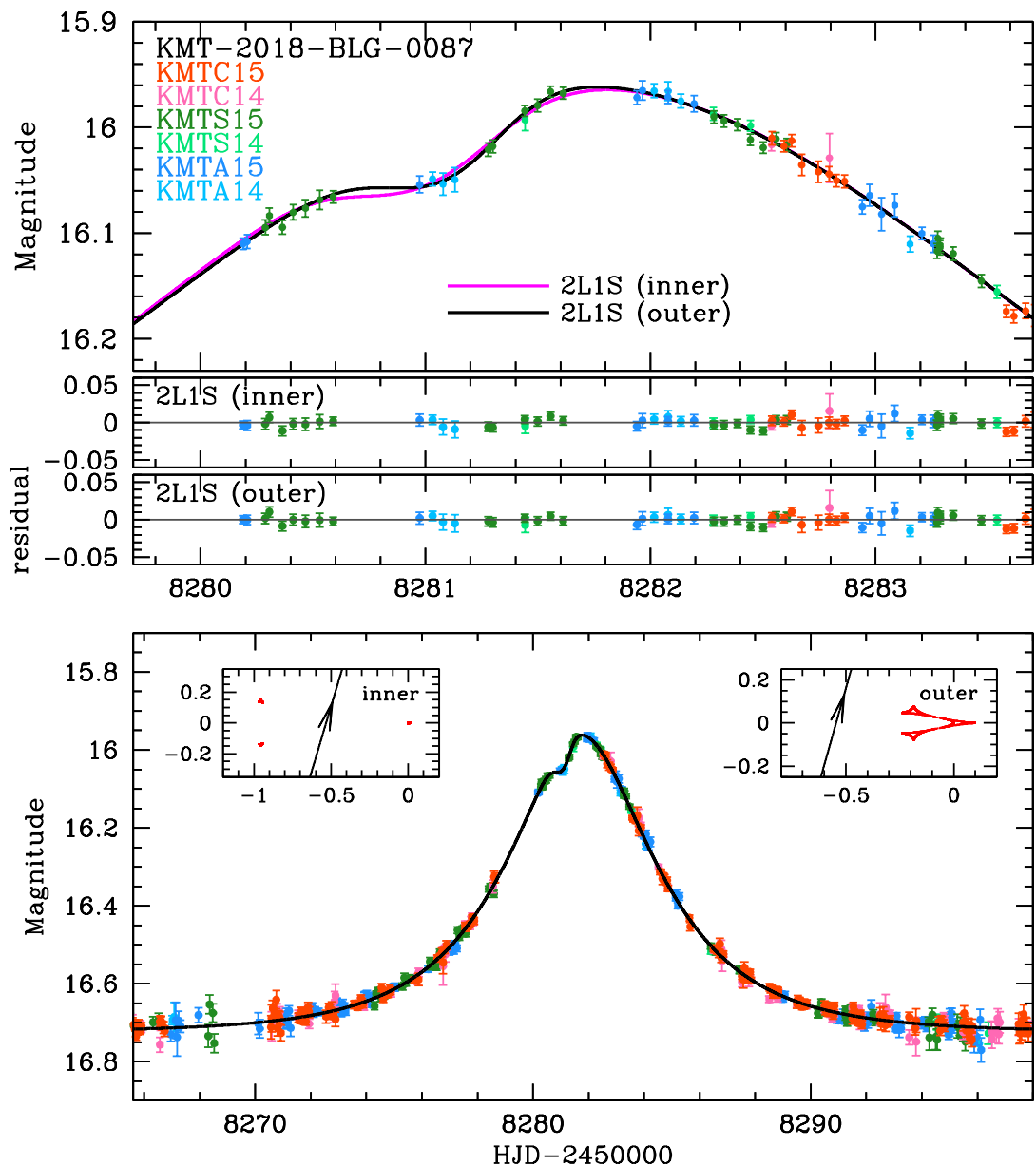}
\caption{Data (color-coded by observatory and field) 
together with the predictions
and residuals for the models of KMT-2018-BLG-0087 specified in 
Table~\ref{tab:kb0087parms}.  The short ``dip'' at $t_{\rm anom}= 8281.73$
is caused by the source crossing the trough that runs along the minor 
image axis due to a $\log q\sim -2.6$ super-Jovian mass-ratio planet.  
It is subject to the ``inner/outer'' degeneracy. See insets.
}
\label{fig:0087lc}
\end{figure}

Figure~\ref{fig:0087lc} shows an approximately 1L1S light curve with
\citet{pac86} parameters $t_0=8281.73$, $u_0=0.53$, and $t_\e=4.55\,$days,
punctuated by a short dip at $t_{\rm anom}=8281.1$, i.e., 
$\tau_{\rm anom} =-0.138$.  The dip is featureless, while $u_0\gg |\tau_{\rm anom}|$,
indicating a roughly vertical source trajectory.  Hence, we expect two solutions
(inner/outer degeneracy) with $\alpha = 285^\circ$ and $s^\dagger_- = 0.767$.
The full duration of the dip is $\Delta t_{\rm dip}=1.0\,$days.
Equation~(\ref{eqn:qeval}) then predicts $q=3.9\times 10^{-3}$.

The grid search indeed returns two solutions, whose refinement leads to the
parameters given in Table~\ref{tab:kb0087parms}.  
These precisely confirm the first two heuristic predictions,
with $\alpha= 286^\circ$ and 
$s^\dagger\equiv \sqrt{s_{\rm inner}s_{\rm outer}}=0.757$, while (as is often the
case, \citealt{kb190253}), the mass-ratio prediction is only qualitatively 
confirmed.  The outer solution
is significantly (but not overwhelmingly) favored at $\Delta\chi^2=5.3$.

This is another super-Jovian mass-ratio planet, with $\log q\simeq -2.65$.

While the constraint on $\rho$, i.e., 
$\rho<0.110$ or $\rho<0.096$ (at $2.5\,\sigma$) is very similar
to KMT-2018-BLG-0030 (and, as we will see in Section~\ref{sec:cmd}, the
source sizes are also similar), this constraint will, in contrast to that
case,  ultimately prove
to be significant.  This is because the Einstein timescale is much shorter.
Indeed, at $t_\e\sim 4.55\,$days, KMT-2018-BLG-0087 is one of the shortest
bound-planet events yet detected \citep{kb162605}.  As we will discuss in
Sections~\ref{sec:cmd-kb180087} and \ref{sec:phys-kb180087}, these 
characteristics imply that the host is most likely a low-mass star 
(or possibly brown dwarf) in the bulge.

Although the anomaly is well-fitted by a minor-image ``dip'' and
therefore is not expected to be compatible with a 1L2S model, we nevertheless
check this possibility as a matter of due diligence.  As anticipated, we
find that 1L2S is ruled out, with 
$\Delta\chi^2=\chi^2({\rm 1L2S}) - \chi^2({\rm 2L1S})=100.1$.

Due to the extreme shortness of the event, we do not attempt to measure
the microlens parallax, $\bpi_\e$.

Of minor note, KMT-2018-BLG-0087 lies in the very small region of
overlap between KMT fields BLG14 and BLG15 and therefore is a
rare case of a sub-prime-field event with a $\Gamma=2\,{\rm hr}^{-1}$ cadence.

\subsection{{KMT-2018-BLG-0247} 
\label{sec:anal-kb180247}}

\begin{figure}
\epsscale{1.0}
\plotone{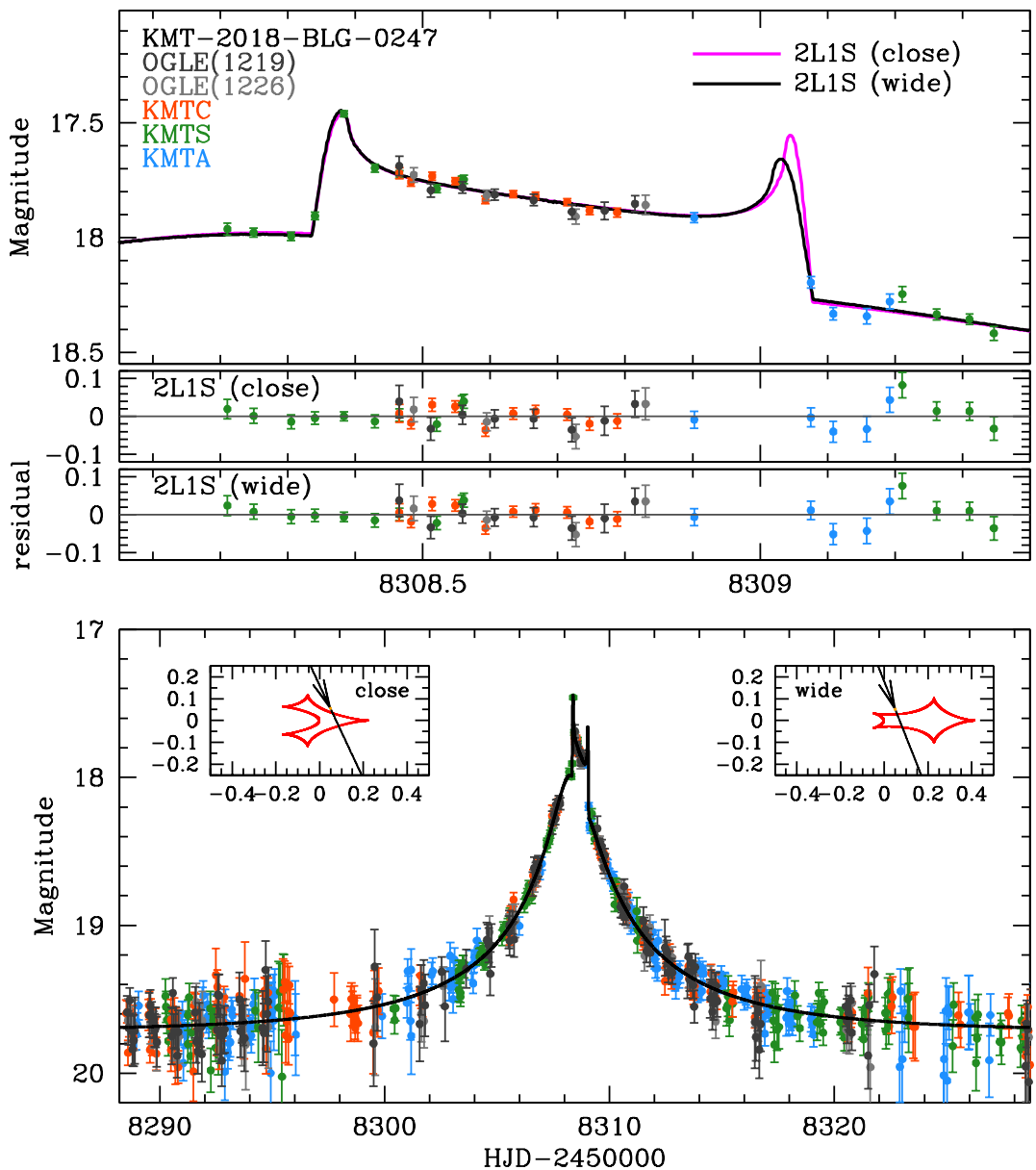}
\caption{Data (color-coded by observatory and field) 
together with the predictions
and residuals for the models of KMT-2018-BLG-0247 specified in 
Table~\ref{tab:kb0247parms}.  The ``double-horned profile'' centered
at $t_{\rm anom}= 8305.70$
is caused by the source crossing the main body of a resonant caustic
due to a $\log q\sim -2.2$ super-Jovian mass-ratio planet.  
It is subject to a ``close/wide'' degeneracy (see insets) due to the
lack of data on the caustic exit..
}
\label{fig:0247lc}
\end{figure}

Figure~\ref{fig:0247lc} shows a short ($\sim 0.7\,$day) double horned
profile centered at $t_{\rm anom}=8305.70$ just after the peak of a normal
1L1S event with parameters $t_0=8308.42$, $u_0=0.065$, and $t_\e=10.6\,$days,
yielding $\tau_{\rm anom}=0.0264$, $u_{\rm anom} = 0.070$ and so
$s_+^\dagger = 1.036$ and $\alpha=68^\circ$.

The grid search returns two solutions, whose refinements are given in 
Table~\ref{tab:kb0247parms}
and whose geometries are shown in Figure~\ref{fig:0247lc}.  Note that both
geometries have the source crossing the neck of a resonant caustic, which
does not appear to be directly related to either of the degeneracies
(inner/outer of \citealt{gaudi97} or close/wide of \citealt{griest98})
that were predicted in advance and that were unified in 
Equation~(\ref{eqn:heuristic}) by \citet{kb211391} following the conjecture of
\citet{ob190960}.  Subsequently, \citet{zhang22} investigated the
origins of unification at the level of the lens equation.
Nevertheless, the two solutions combine to
yield $s^\dagger\equiv\sqrt{s_+ s_-} = 1.043$, in excellent agreement with the
prediction.

Both solutions imply another super-Jovian mass-ratio planet, $\log q = -2.2$.

Figure~\ref{fig:0247lc} shows that only the caustic entrance is resolved.
The fact that the source enters the caustic at different angles
in the two solutions leads to different values of $\rho$, which
are proportional to the cosine of this angle.  Note that if the exit
had been resolved by the data, then the degeneracy would have been
broken.  In this sense, it is accidental.

Due to the short $t_\e$, we do not attempt to measure the microlens parallax,
$\bpi_\e$.

KMT-2018-BLG-0247 is one of three previously known planets that
are analyzed here for the first time.

\subsection{{OGLE-2018-BLG-0298}
\label{sec:anal-ob180298}} 

\begin{figure}
\epsscale{1.0}
\plotone{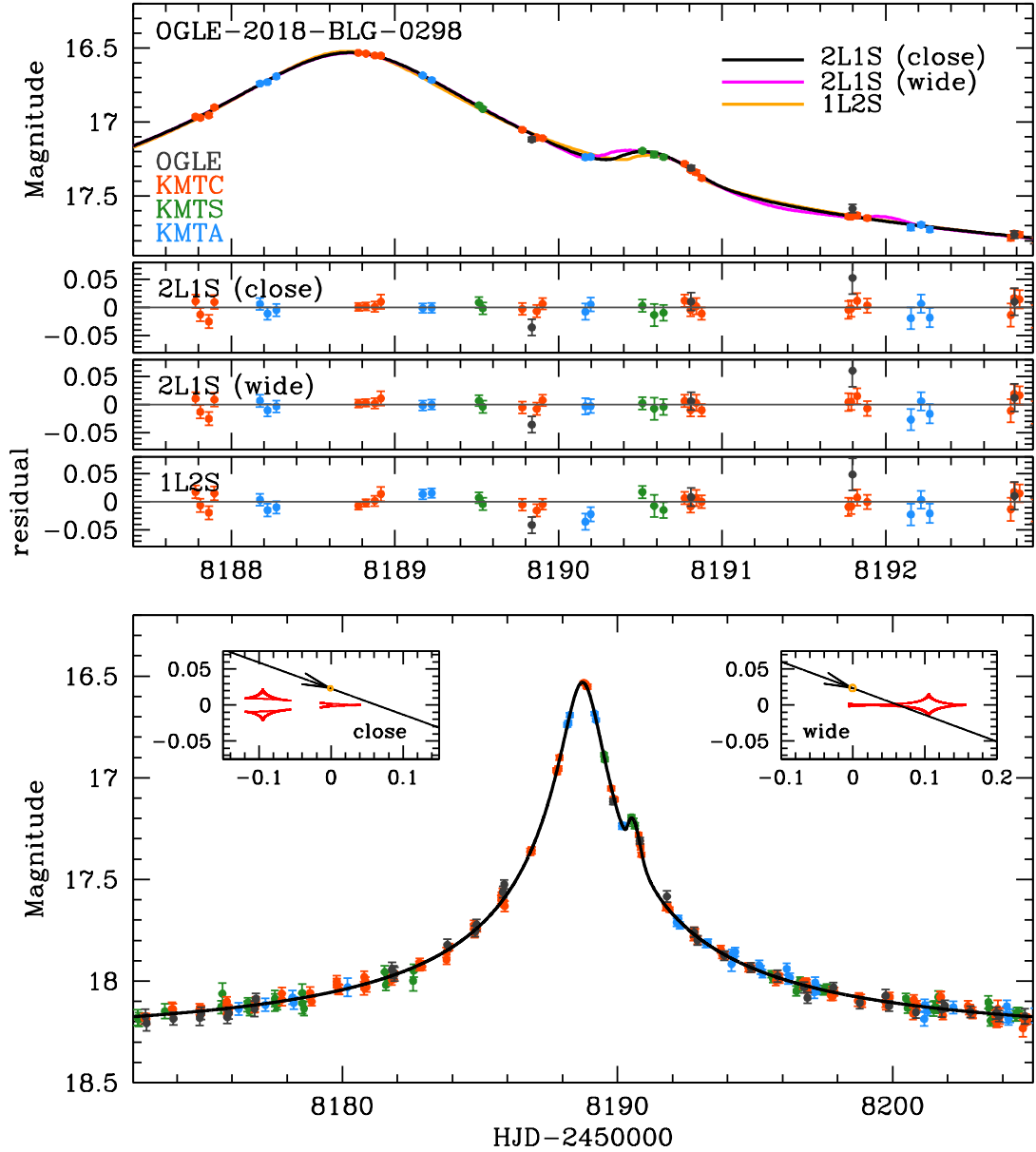}
\caption{Data (color-coded by observatory) together with the predictions
and residuals for the models of OGLE-2018-BLG-0298 specified in 
Table~\ref{tab:ob0298parms}.  The short ``bump'' 
at $t_{\rm anom}= 8190.6$
is caused by the source crossing a ridge in or extending from the 
major-image side of the caustic structure 
due to a $\log q\sim -3.7$ sub-Saturn mass-ratio planet.  
It is subject to a ``close/wide'' degeneracy.  See insets.
The 1L2S model is disfavored by $\Delta\chi^2=33.7$ and, in addition,
is heavily disfavored by kinematic arguments.  Hence, it is excluded.
}
\label{fig:0298lc}
\end{figure}

Figure~\ref{fig:0298lc} shows a brief bump on the falling wing
at $t_{\rm anom}=8190.6$ of a normal
1L1S event with parameters $t_0=8188.74$, $u_0=0.021$, and $t_\e=32\,$days,
yielding $\tau_{\rm anom}=0.058$, $u_{\rm anom} = 0.062$, and so
$s_+^\dagger = 1.031$ and $\alpha=20^\circ$.

The grid search returns two solutions, whose refinements are shown in
Table~\ref{tab:ob0298parms}.  
Based on caustic geometries shown in Figure~\ref{fig:0298lc},
these might plausibly be identified as a case of the ``inner/outer'' degeneracy,
although
this would be very far from the original conception of \citet{gaudi97}.  Even
so, the parameters $s^\dagger\equiv \sqrt{s_+ s_-}= 1.016$ and $\alpha=20^\circ$
are in reasonably good agreement with the heuristic prediction.
Note that the wide solution has a cusp approach at about 8192.5, which is
also preceded by a dip.  However, 
these putative features are not probed by any data. See Figure~\ref{fig:0298lc}.
In this sense, the degeneracy is accidental.

Because the bump is featureless, we check for 1L2S solutions, but we find
that these are excluded at $\Delta\chi^2=33.7$.  In addition, the solution
gives $t_{*,2} = \rho_2 t_\e = 0.21\,$days and $q_F=0.013$, indicating
that the second source lies about 7 mag below the clump
(see Section~\ref{sec:cmd-ob180298}), and so implying
$\theta_{*,2}\sim 0.3\,\muas$.  These yield 
$\mu_\rel=\theta_{*,2}/t_{*,2}=0.55\,\masyr$, which is quite improbable
according to Equation~(\ref{eqn:probmu}).
Therefore, we consider the 1L2S models to be decisively excluded.

Given the intermediate timescale, $t_\e\sim 32\,$days, and moderately
faint source, $I_S\sim 20$, it would be unlikely
that a full, 2-dimensional (2-D) parallax could be measured.
Nevertheless, a 1-D parallax measurement is plausible.  For example,
\citep{2018prime} found two such cases.  See their Figure~3.
Indeed, for the four cases, 
(close $u_0>0$,
wide $u_0>0$,
close $u_0<0$,
wide $u_0<0$)
we find four, very similar, highly-elongated error ellipses, with
$(\pi_{\e,\parallel},\pi_{\e,\perp},\psi) =
 (0.00\pm 0.11, -0.08\pm 0.63, 272.1^\circ)$,
$(-0.02\pm 0.11, -0.06\pm 0.68, 272.0^\circ)$,
$(0.00\pm 0.11, -0.19\pm 0.64, 271.2^\circ)$, and
$(-0.02\pm 0.11, -0.24\pm 0.64, 271.5^\circ)$, respectively\footnote{
See Figure 3 of \citet{mb03037} for sign conventions, keeping in mind that
OGLE-2018-BLG-0298 peaked before opposition while MOA-2003-BLG-037
peaked after opposition.}.
Here $\pi_{\e,\parallel}$, and $\pi_{\e,\perp}$ are the minor and major axes
of the error ellipse, which are so named because their orientation
angle $\psi$ (north through east) is approximately parallel to the
projected position of the Sun at the peak of the event.  In all cases,
the improvement from adding parallax (and orbital motion, 
see Section~\ref{sec:anal-preamble}) is $\Delta\chi^2<1$ for four degrees
of freedom.  We find that in all four cases, the numerical contours
are well described as Gaussian realizations of the 3-parameter representations
listed above.

Nevertheless, the constraint implied by this parallax measurement
is moderately significant.  As discussed by \citet{ob150479}, we are not
trying to ``detect'' parallax: it is known a priori that 
$\pi_\e\equiv \pi_\rel/\theta_\e$ is strictly non-zero.  Hence, limits,
even 1-D limits, on the parallax vector can be constraining when
combined with prior information from a Galactic model.  We discuss
the implementation of these constraints in Section~\ref{sec:phys-ob180298}.
Here, we only remark that the remaining parameters change by much
less than their error bars, and they would change even less
if we were to apply Galactic-model priors to the parallax.  Therefore,
we report the static-model solutions in Table~\ref{tab:ob0298parms}.

This is a sub-Saturn mass-ratio planet, $\log q= -3.7$, the only
securely planetary event in this paper to lie in this range.
By comparison, among 2018 prime-field unambiguously planetary events
that were newly discovered by AnomalyFinder, there were 5 with
$q< 2\times 10^{-4}$, namely, 
OGLE-2018-BLG-0383 \citep{ob180383},
OGLE-2018-BLG-0506,
OGLE-2018-BLG-0516, and
OGLE-2018-BLG-0977 \citep{kb190253}, and
OGLE-2018-BLG-1126 \citep{2018prime}.

\subsection{{KMT-2018-BLG-2602} 
\label{sec:anal-kb182602}}

\begin{figure}
\epsscale{1.0}
\plotone{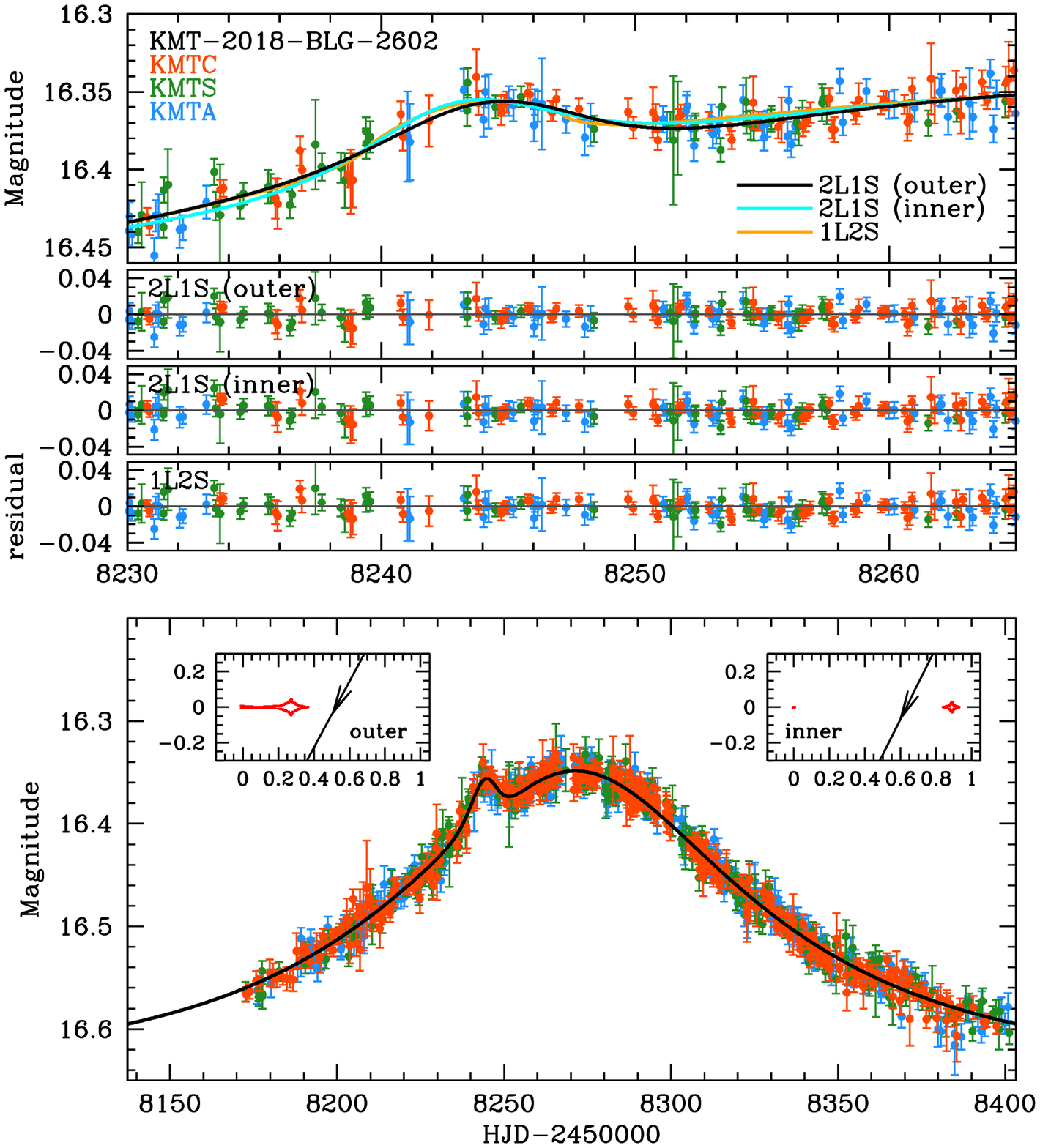}
\caption{Data (color-coded by observatory) together with the predictions
and residuals for the models of KMT-2018-BLG-2602 specified in 
Table~\ref{tab:kb2602parms}.  The short ``bump'' 
at $t_{\rm anom}= 8243.8$
is caused by the source crossing a ridge extending from the 
planetary caustic (or planetary wing of the resonant caustic)
due to a $\log q\sim -2.8$ Jovian mass-ratio planet.  
It is nominally subject to a ``inner/outer'' degeneracy  (see insets), but
we adopt the ``outer'' solution because it is favored by $\Delta\chi^2=10.3$.
The 1L2S model is disfavored by $\Delta\chi^2=30.7$ and, in addition,
is very heavily disfavored by kinematic arguments.  Hence, it is excluded.
}
\label{fig:2602lc}
\end{figure}

Figure~\ref{fig:2602lc} shows an approximately 1L1S light curve with
parameters $t_0=8270.3$, $u_0=0.51$, and $t_\e=98\,$days, punctuated by 
a short, smooth bump at $t_{\rm anom} = 8243.8$, yielding 
$\tau_{\rm anom} = -0.27$ and $u_{\rm anom}=0.58$, which imply
$s_+^\dagger = 1.33$ and $\alpha=118^\circ$.

The grid search returns two solutions, whose refined parameters are shown
in Table~\ref{tab:kb2602parms}.  
These agree with the heuristic predictions.  In particular
$s^\dagger \equiv \sqrt{s_+ s_-} = 1.35$.  The close solution is favored
by $\Delta\chi^2=10.4$, and we therefore choose it for our adopted
solution.  However, the mass ratios of two solutions are nearly identical:
this is Jovian mass-ratio planet: $\log q =-2.8$.
Figure~\ref{fig:2602lc} shows that the two caustic topologies
are related by an ``inner/outer'' degeneracy in which (as is often the case),
the ``outer'' topology has a resonant caustic.

Because the bump is featureless, it could in principle be generated
by a second source, rather than a second lens.  However, we find
that 1L2S solutions are excluded at $\Delta\chi^2=30.7$.  Moreover,
from Table~\ref{tab:kb2602parms}, 
we see that the second-source self-crossing time is
$t_{*,2}\sim 3.9\,$days. while the large flux ratio would imply that
the second source lies about 7.7 mag below the clump, indicating
a second-source radius $\theta_{*,2}\sim 0.3\,\muas$.  Combined, these
would imply $\mu_\rel=\theta_{*,2}/t_{*,2} = 0.028\,\masyr$, which
is extraordinarily unlikely.  See Equation~(\ref{eqn:probmu}).

By contrast, in the 2L1S solutions, there are only upper limits
on $\rho$, so no such issues arise.

In spite of the long duration of the event, we do not attempt a parallax
analysis because the baseline cannot be properly constrained within the
2018 season, and we find small photometric offsets between seasons
that would render multi-season analysis questionable.

KMT-2018-BLG-2602 is one of three previously known planets that
are analyzed here for the first time.

\subsection{{OGLE-2018-BLG-1119}
\label{sec:anal-ob181119}} 

\begin{figure}
\epsscale{1.0}
\plotone{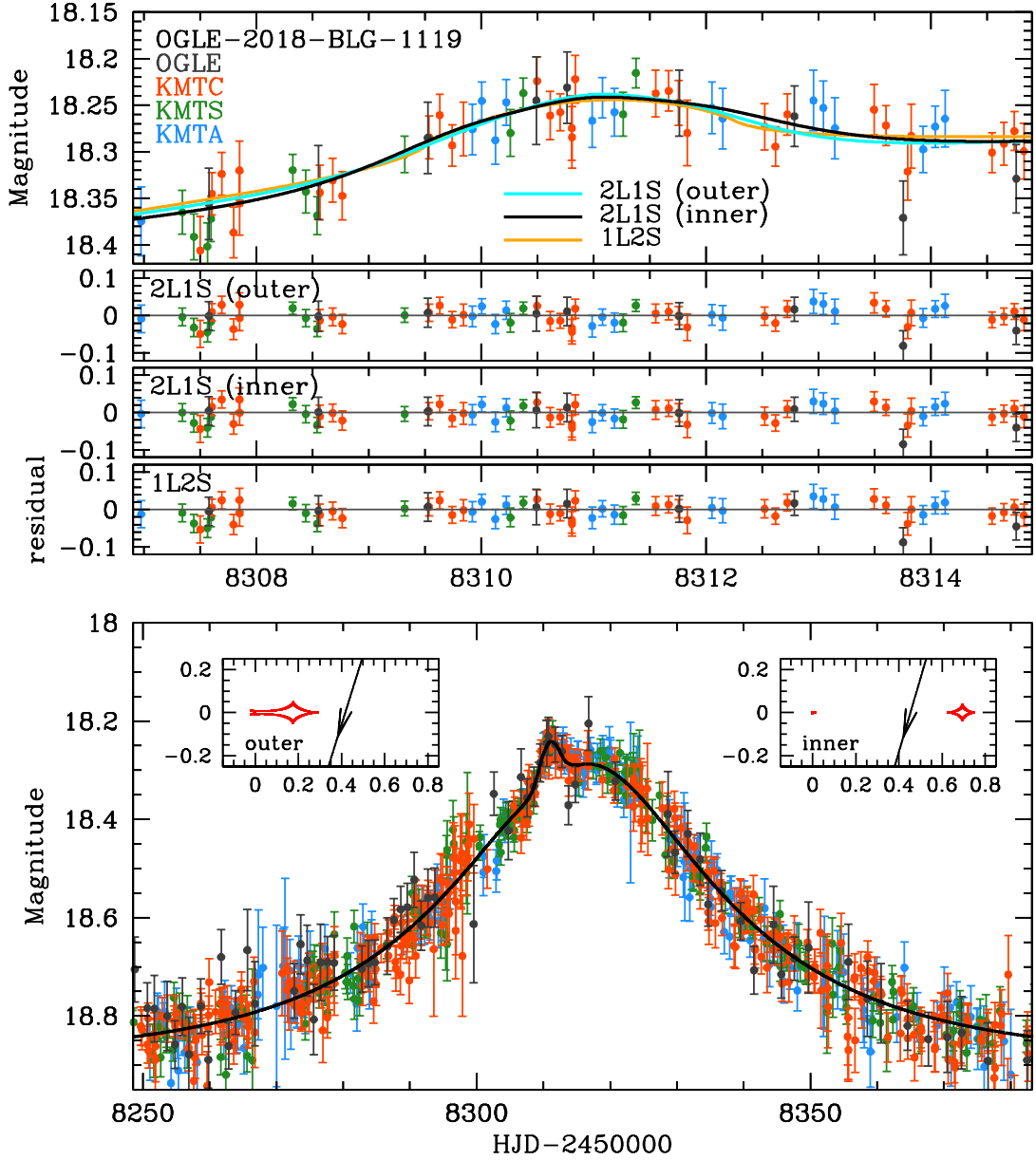}
\caption{Data (color-coded by observatory) together with the predictions
and residuals for the models of OGLE-2018-BLG-1119 specified in 
Table~\ref{tab:ob1119parms}.  The short ``bump'' 
at $t_{\rm anom}= 8310.7$
is caused by the source crossing a ridge extending from the 
planetary caustic (or planetary wing of the resonant caustic)
due to a $\log q\sim -2.75$ Jovian mass-ratio planet.  
It is subject to a ``inner/outer'' degeneracy. See insets.
The 1L2S model is disfavored by $\Delta\chi^2=14.0$ and, in addition,
is very heavily disfavored by kinematic arguments.  Hence, it is excluded.
}
\label{fig:1119lc}
\end{figure}

Figure~\ref{fig:1119lc} shows an approximately 1L1S light curve with
parameters, $t_0=8316.0$, $u_0=0.43$, and $t_\e=40\,$days, 
punctuated by a small bump before the peak at $t_{\rm anom}=8310.7$, i.e., with
$\tau_{\rm anom} = -0.133$, and so $u_{\rm anom} = 0.45$.  These values predict
$s_+^\dagger = 1.25$ and $\alpha=106^\circ$.

The grid search yields two solutions, whose refinements are shown in
Table~\ref{tab:ob1119parms}.  
They confirm $s^\dagger \equiv \sqrt{s_+ s_-} = 1.24$ and 
$\alpha=107^\circ$.  Figure~\ref{fig:1119lc} shows that this is an
inner/outer degeneracy in which the outer solution has a resonant caustic.

Given the featureless character of the bump (which, in the 2L1S models
is explained by a ridge crossing), we also fit the event with a 1L2S model.
Table~\ref{tab:ob1119parms}
shows that this is disfavored by $\Delta\chi^2=14.0$.  In itself,
this argues strongly against the 1L2S hypothesis, but does not
completely exclude it.  

We therefore also investigate the physical plausibility of the 1L2S solution.
Table~\ref{tab:ob1119parms} 
shows that $t_{*,2}\equiv\rho_2t_\e=1.40\pm 0.27\,$days 
and $q_F=4.2\times 10^{-3}$.
We will see in Section~\ref{sec:cmd-ob181119} that the latter implies that
the second source lies $\sim 10\,$ mag below the clump, and hence it has
$\theta_{*,2}\sim 0.15\,\muas$, implying 
$\mu_\rel=\theta_{*,2}/t_{*,2}\sim 0.039\,\masyr$.
This is extraordinarily improbable, i.e., $p=2\times 10^{-7}$, according
to Equation~(\ref{eqn:probmu}).  Hence, we consider the 1L2S solution
to be excluded by the combination of $\Delta\chi^2$ and kinematic arguments.

This is another Jovian-class mass-ratio planet, $\log q=-2.75$.

Due to the event's low amplitude and faint source, we do not
attempt a microlens parallax analysis.

\subsection{{KMT-2018-BLG-0173} 
\label{sec:anal-kb180173}}

\begin{figure}
\epsscale{1.0}
\plotone{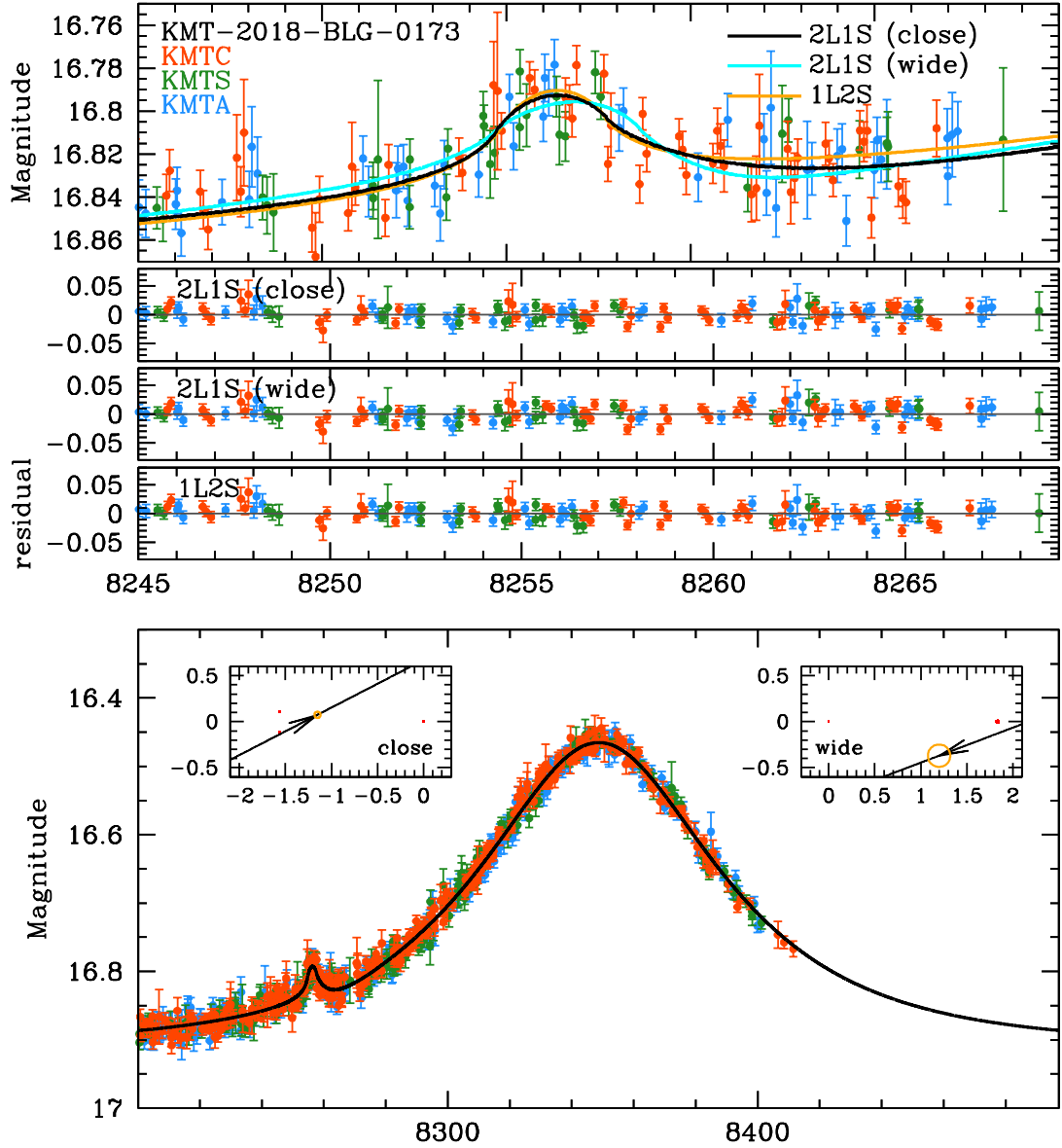}
\caption{Data (color-coded by observatory) together with the predictions
and residuals for the models of KMT-2018-BLG-0173 specified in 
Table~\ref{tab:kb0173parms}.  The short ``bump'' 
at $t_{\rm anom}= 8256$ has 3 possible explanations: source hitting
minor image caustic (left inset), source hitting major image caustic
(right inset) [both with $\log q=-3.0$], or 1L2S model.
The last cannot be decisively rejected based on current data
(see Sections~\ref{sec:anal-kb180173} and \ref{sec:cmd-kb180173}).
Therefore, this event should not be cataloged as planetary.
}
\label{fig:0173lc}
\end{figure}

Figure~\ref{fig:0173lc} shows an approximately 1L1S light curve with
parameters of (assuming zero blending)
$t_0=8348.7$, $u_0=0.79$, and $t_\e=52\,$days, punctuated by a small bump
far out on the leading wing at $t_{\rm anom}=8256$, i.e., with
$\tau_{\rm anom} = -1.78$ and so $u_{\rm anom} = 1.95$.  At this separation,
the planetary caustics are generally small and weak.  Hence, it is more
likely that the featureless bump is due to the source at least partially
enveloping a planetary
caustic (either major or minor image), rather than crossing the ridge associated
with a major-image caustic.  We therefore report both branches of $s^\dagger_\pm$,
i.e., $s_-^\dagger = 0.42$ and $s_+^\dagger = 2.37$, with corresponding values
of $\alpha_-= 156^\circ$ and $\alpha_+ = 336^\circ$.

The grid search indeed returns two solutions, whose refinements are 
shown in Table~\ref{tab:kb0173parms}, 
and which are roughly in accord with these predictions. 
Note that the heuristic formalism naively predicts a second wide
solution at $s_{\rm wide,2} = (s^\dagger)^2/s_{\rm wide} = 2.43$.  This would
basically correspond to the source enveloping the caustic from its
left side (as opposed to the right-side envelopment shown in 
Figure~\ref{fig:0173lc}).  However, the error in $s$ from
Table~\ref{tab:kb0173parms} already essentially covers this alternative
solution at $1.5\,\sigma$, 
and we find that seeding an MCMC with this solution leads to
a convergence at the reported solution.  Hence, there is a relatively large
continuous degeneracy in $s$, rather than a discrete degeneracy.

Both solutions have Jovian-class mass ratios, $\log q =-3.0$.
See Table~\ref{tab:kb0173parms}.

Two arguments favor the close solution.  First, $\Delta\chi^2=11.2$,
which is significant but not overwhelming evidence.  Second,
while the close solution is consistent with small $\rho$ (in particular,
$\rho=0$ is disfavored by only $\Delta\chi^2=3.5$), the wide solution
is well localized at $\rho_{\rm wide}\sim 0.125$, corresponding to 
$t_{*,\rm wide}\sim 6.5\,$days.  We will show in 
Section~\ref{sec:cmd-kb180173} that for the wide solution 
$\theta_{*,\rm wide}\sim 6.0\,\muas$, which would imply 
$\mu_{\rel,\rm wide}=\theta_{*,\rm wide}/t_{*,\rm wide}\sim 0.34\,\masyr$.
According to Equation~(\ref{eqn:probmu}), this has a probability
$p\sim 1.5\times 10^{-4}$.  (Alternatively, if we force $\rho=0$,
the best-fit solution has an additional $\Delta\chi^2=18.0$, which
would be formally disfavored by a similar factor,
$p=\exp(-(18.0-3.5)/2)=7\times 10^{-4}$.)\ \
The combination of these two arguments
leads us to strongly favor the close over the wide solution.

However, because the anomaly is a featureless bump, we must 
also check the 1L2S model.  See Table~\ref{tab:kb0173parms}.  This model
is disfavored by three different arguments.  First,
it is disfavored by $\Delta\chi^2=10.8$, i.e., very
similar to the difference between the close and wide models.  If Gaussian
statistics applied, this would correspond to $p = 4.5\times 10^{-3}$,
which would effectively settle the matter.  However, this $\chi^2$ difference
derives from subtle differences in the models over several weeks 
(see Figure~\ref{fig:0173lc}), and so could be impacted by equally
subtle long-term systematics, which would be difficult to track down.
Hence, we look for additional evidence.  

The second argument is that the best-fit color of the second source
is essentially identical to that of the first source,
$\Delta(V-I) \equiv (V-I)_2 - (V-I)_1\simeq 0$, 
whereas its magnitude is $-2.5\log(q_F)=7.5$
mag fainter, and hence should be substantially redder, $\Delta(V-I)\sim 1$.
See Section~\ref{sec:cmd-kb180173}.  That is, if the 2L1S model is correct,
the amplitude of the ``bump'' should be essentially identical in $V$ and $I$,
while for the 1L2S model, it should be suppressed by a factor
$10^{-0.4\Delta(V-I)}\sim 0.4$.  There are 6 $V$-band points (2 from each 
observatory) over portions of the bump that deviate from the 1L1S model
by at least of order the $V$-band error bars ($\sim 0.03\,$mag), which
track the $I$-band points (not shown).  Unfortunately, this is a 
weak test because of the paucity of $V$-band
data and the relatively large error bars.
We find that if we enforce $\Delta(V-I)=1$, then $\chi^2$ is only increased
by $\Delta\chi^2=2.5$.  Nevertheless, in contrast to the $\Delta\chi^2=10.8$ 
difference from the $I$-band fit, this determination is not subject
to potential systematic errors from long-term trends in the light curve:
it is a purely differential measurement from $V$ and $I$ measurements
taken under essentially identical conditions, just 2 minutes apart.

Third, a kinematic argument (similar to the one given above against
the 2L1S wide solution) further argues against the 1L2S
solution.  The flux ratio, $q_F\sim 10^{-3}$ would imply
$\theta_{*,2}\sim 0.3\,\muas$ for the radius of the second source.
See Section~\ref{sec:cmd-kb180173}.
On the other hand, $t_{*,2}\sim 2.0\,$days, implying 
$\mu_\rel\sim \theta_{*,2}/t_{*,2}=0.055\,\masyr$.  According to 
Equation~(\ref{eqn:probmu}), this would have probability $p<10^{-6}$.

Nevertheless, we note that there is a partial loophole to this argument.
While the MCMC is well localized near $\rho_2\sim 0.04$, solutions
with $\rho_2\sim 0$ are excluded at only (an additional) $\Delta\chi^2=6.5$.
Thus, a strict comparison of the 2L1S (close) solution and the 
1L2S ($\rho_2=0$) solution, in effect, favors the former by 
$\Delta\chi^2\sim 10.8+6.5-3.5=14.8$.  Moreover, if we combine
this preference, which is subject to effects of long-term systematics,
with the source-color argument above, which does not, we obtain
$\Delta\chi^2\sim 14.8 + 2.5=16.3$.  

We defer weighing these various pieces of evidence
until Section~\ref{sec:cmd-kb180173},
where we will bring to bear additional information of the source-star
characteristics in these three different models.


\subsection{{KMT-2018-BLG-1497}
\label{sec:anal-kb181497}}

\begin{figure}
\epsscale{1.0}
\plotone{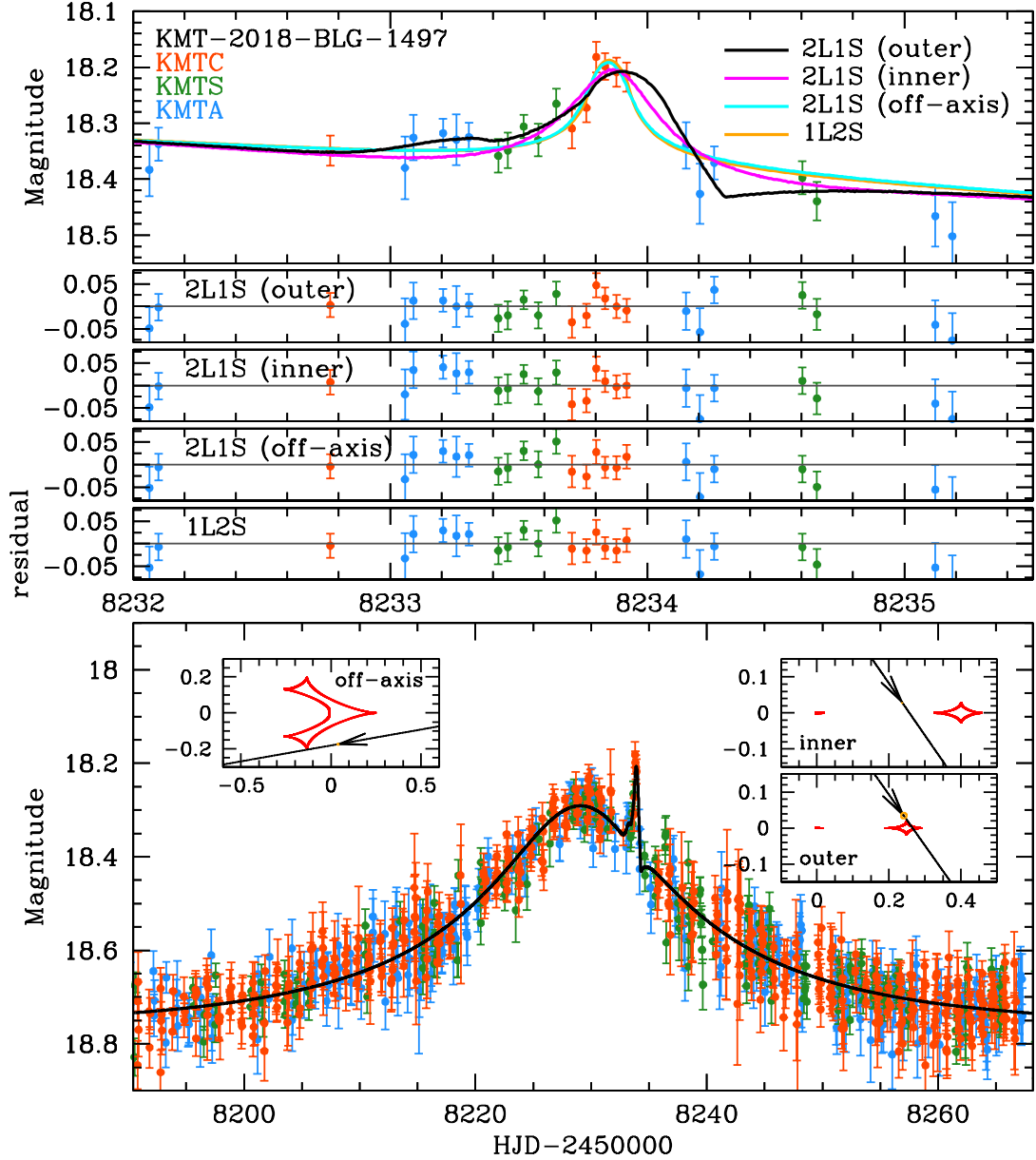}
\caption{Data (color-coded by observatory) together with the predictions
and residuals for the models of KMT-2018-BLG-1497 specified in 
Table~\ref{tab:kb1497parms}.  The short ``bump'' 
at $t_{\rm anom}= 8233.9$ has 4 possible explanations: 2 from an
``inner/outer'' degeneracy (right insets), one from an off-axis cusp
approach (left inset) [together with $-3.7\la\log q\la-1.8$], or 1L2S model.
The last has $\Delta\chi^2=2.1$, and there are no other arguments against it.
Therefore, this event should not be cataloged as planetary.
}
\label{fig:1497lc}
\end{figure}

Figure~\ref{fig:1497lc} shows an approximately 1L1S light curve with
parameters, $t_0=8229.1$, $u_0=0.21$, and $t_\e=31\,$days, with a short,
possibly structured bump at $t_{\rm anom} = 8233.9$, yielding 
$\tau_{\rm anom} = +0.155$ and $u_{\rm anom}=0.26$, which imply
$s_+^\dagger = 1.14$ and $\alpha=59^\circ$.  

The grid search returns 3 solutions, whose refinements are shown in
Table~\ref{tab:kb1497parms}.  
Two of these approximately correspond to the heuristic
prediction, with $s^\dagger =\sqrt{s_+ s_-}=1.17$ and $\alpha=55^\circ$.
Although the discrepancies are modest in absolute terms, they are
significantly larger than is typically the case.  Figure~\ref{fig:1119lc}
shows that while this is an inner/outer degeneracy, the outer solution
has a caustic crossing (which is favored by the early KMTA points), while
the inner solution does not.  Hence, we do not expect the heuristic
formalism to work perfectly.

More importantly, there is a third solution, in which the bump is
due to an off-axis cusp approach.  See Figure~\ref{fig:1119lc}.
Table~\ref{tab:kb1497parms} shows that the 3 solutions cover a factor 80 range
of mass ratios, $q$, while the full range of $\chi^2$ is only
$\Delta\chi^2=4.2$.  Hence, even if this could be confidently accepted
as a planetary event, the planet's characteristics would be extremely
uncertain.

Moreover, we find that there is a 1L2S model with $\Delta\chi^2=2.1$.
See Table~\ref{tab:kb1497parms}.
Because $\rho_2$ is not confidently measured for this solution,
we cannot make kinematic arguments against it.

Therefore, first, we cannot be confident that this is a planetary event,
and second, even if it is, we cannot determine the planet's mass ratio.
Therefore, this event should not be cataloged as ``planetary'', and,
in particular, it should not enter mass-ratio studies.

Due to the ambiguous nature of the event, we do not attempt a microlens parallax
analysis.

\subsection{{KMT-2018-BLG-1714}
\label{sec:anal-kb181714}}

\begin{figure}
\epsscale{1.0}
\plotone{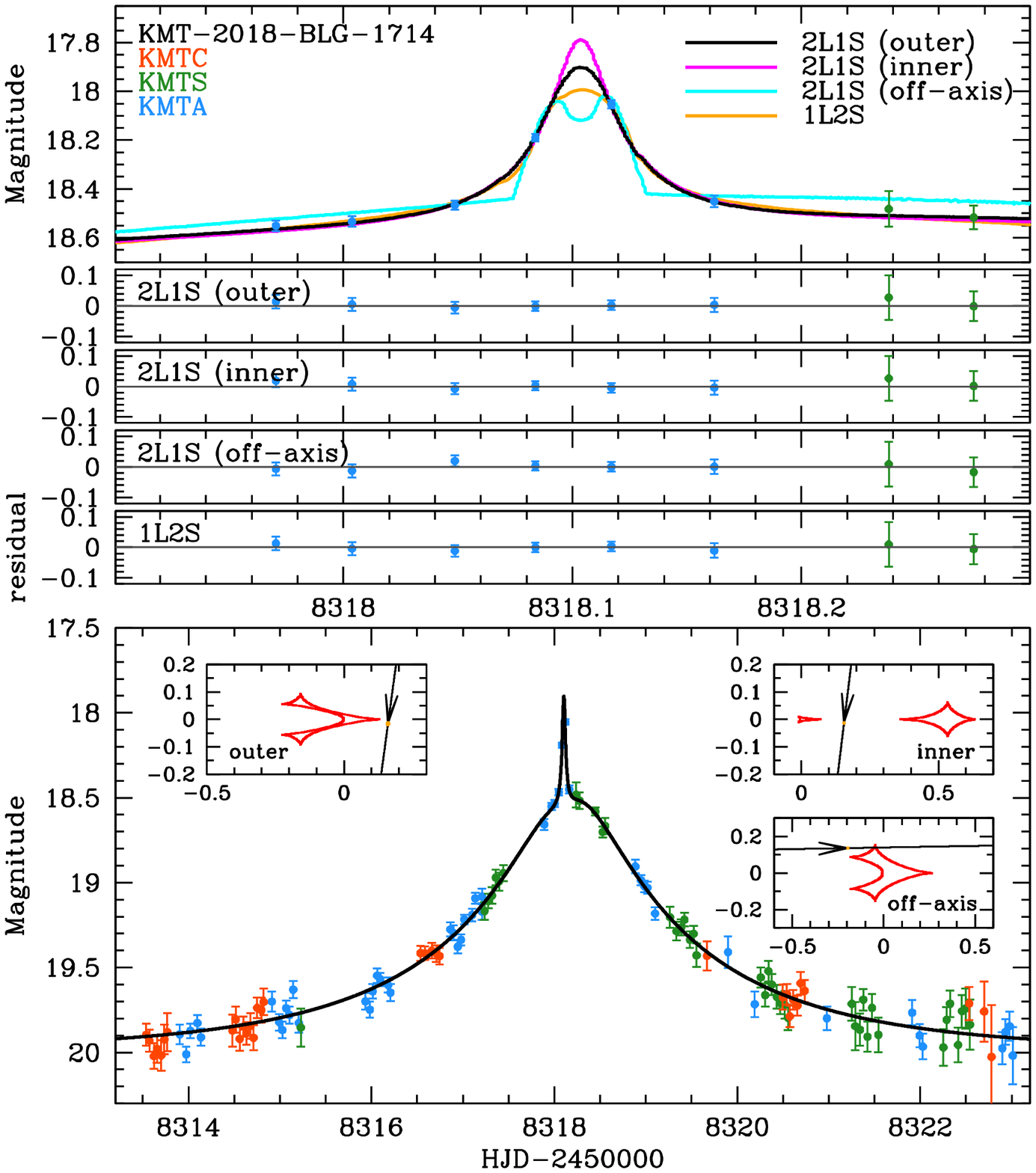}
\caption{Data (color-coded by observatory) together with the predictions
and residuals for the models of KMT-2018-BLG-1714 specified in 
Table~\ref{tab:kb1714parms}.  The short ``bump'' 
at $t_{\rm anom}= 8318.10$ has 4 possible explanations: 2 from an
``inner/outer'' degeneracy (top insets), one from an off-axis cusp
approach (lower right inset) [together with $-2.4\la\log q\la-1.8$], 
or 1L2S model.
The last has $\Delta\chi^2=0.9$, and there are no other arguments against it.
Therefore, this event should not be cataloged as planetary.
}
\label{fig:1714lc}
\end{figure}

Figure~\ref{fig:1714lc} shows an approximately 1L1S light curve with
parameters $t_0=8318.17$, $u_0=0.16$, and $t_\e=3.2\,$days, with a short
bump defined by just two points at $t_{\rm anom} = 8318.10$, yielding 
$\tau_{\rm anom} = -0.022$ and $u_{\rm anom}=0.16$, which imply
$s_+^\dagger = 1.08$ and $\alpha=98^\circ$.  The two points were taken
in $\sim 1.5^{\prime\prime}$ seeing, which is good for KMTA, and also
very similar to the previous 4 points.  Similarly, the background
was low and steady.  There is nothing unusual about these points,
so we conclude that the increase in recorded flux is of astrophysical
origin.

However, a grid search returns three solutions, rather than the usual
one or two.  Two of these solutions constitute
an inner/outer pair, with similar $\alpha=97^\circ$ and
$s^\dagger\equiv \sqrt{s_+ s_-} = 1.09$ to those anticipated from the
heuristic analysis (see Table~\ref{tab:kb1714parms}).  
However, the third solution has a completely
different topology, in which the bump is generated by an off-axis
cusp of a resonant caustic.  See Figure~\ref{fig:1714lc}.  The mass
ratio of this solution is about 3.5 times larger than those of the
other two.  While this solution is disfavored by $\Delta\chi^2=7.5$,
it cannot be definitively excluded on these grounds.

Moreover, such a poorly traced bump could be due to an extra source rather
than an extra lens. We find a 1L2S solution with $\Delta\chi^2=0.9$.
Because $\rho_2$ is poorly constrained in this solution, we cannot
develop arguments against it based on physical considerations,
as we could in several other cases.  Therefore, we cannot be certain
that the anomaly of this event is due to a planet, and we cannot
uniquely determine the mass ratio even if it is assumed to be a planet.

Hence, we specifically counsel against cataloging this event as
planetary, and, in particular, we advise against it being used in mass-ratio
studies.

Due to the ambiguous nature of the event, we do not attempt a microlens parallax
analysis.

\section{{Source Properties}
\label{sec:cmd}}

If $\rho$ can be measured from the light curve, then 
one can use standard techniques \citep{ob03262} to determine the
angular source radius, $\theta_*$ and so infer
$\theta_\e$ and $\mu_\rel$:
\begin{equation}
\theta_\e = {\theta_*\over \rho}; \qquad
\mu_\rel = {\theta_\e\over t_\e}.
\label{eqn:mu-thetae}
\end{equation}
However, in contrast to the majority of published by-eye discoveries
(but similarly to most of new AnomalyFinder discoveries reported
in \citealt{ob191053,af2,kb190253,2018prime}), most of the planetary
events reported in this paper have only upper limits on $\rho$,
and these limits are mostly not very constraining.  
As discussed by \citet{2018prime}, in these cases,
$\theta_*$ determinations are not likely to be of much use, either now
or in the future.  Nevertheless, the source color and magnitude measurement
that are required inputs for these determinations may be of use in the
interpretation of future high-resolution observations, either by space
telescopes or adaptive optics (AO) on large ground-based telescopes.
Hence, like \citet{2018prime}, we calculate $\theta_*$ in all cases.

Our general approach is
to obtain pyDIA \citep{pydia} reductions of KMT data at 
one (or possibly several) observatory/field combinations.  These
yield the microlensing light curve and field-star photometry on the
same system.  We then determine the source color by regression of the
$V$-band light curve on the $I$-band light curve, and the source magnitudes
in $I$ by regression on the best-fit model.  While 
\citet{2018prime} were able to calibrate these CMDs using published
field star photometry from OGLE-III \citep{oiiicat} or OGLE-II 
\citep{oiicat1,oiicat2,oiicat3}, only 2 of the 9 sub-prime-field events in
this paper are covered by these catalogs.  Hence, for the remaining 7,
we work directly in the KMTC pyDIA magnitude system.  Because the
$\theta_*$ measurements depend only on photometry relative to the clump,
they are unaffected by calibration.  In the current context, calibration
is only needed to interpret limits on lens light.  Where relevant,
we carry out an alternative approach to calibration, as we explicitly
describe.

We then follow the standard method of \citet{ob03262}.  We adopt the
intrinsic color of the clump $(V-I)_{0,\rm cl}= 1.06$ from \citet{bensby13}
and its intrinsic magnitude from Table~1 of \citet{nataf13}.
We obtain 
$[(V-I),I]_{\rm S,0} = [(V-I),I]_{\rm S} + [(V-I),I]_{\rm cl,0} - [(V-I),I]_{\rm cl}$.
We convert from $V/I$ to $V/K$ using the $VIK$ color-color relations of
\citet{bb88} and then derive $\theta_*$ using the
relations of \citet{kervella04a,kervella04b} for giant and dwarf sources,
respectively.  After propagating errors, we
add 5\% in quadrature to account for errors induced by the overall method.
These calculations are shown in Table~\ref{tab:cmd}.  Where there are multiple
solutions, only the one with the lowest $\chi^2$ is shown.  However,
the values of $\theta_*$ can be inferred for the other solutions by noting
the corresponding values of $I_S$ in the event-parameter tables and using
$\theta_*\propto 10^{-I_S/5}$.  In any case, these are usually the same within
the quoted error bars.

Where relevant, we report the astrometric 
offset of the source from the baseline object.

Comments on individual events follow.

\subsection{{KMT-2018-BLG-0030} 
\label{sec:cmd-kb180030}}

As noted in Section~\ref{sec:anal-kb180030}, the free-blending fit is
consistent with zero blending at the $1\,\sigma$ level, and we therefore
enforced zero blending in the fit.  We also find that  the source
position lies $<15\,\mas$ from the baseline object (i.e., within the 
measurement precision), which implies that if there is any blended
light, it is almost certainly associated with the event, i.e., from
the lens itself, a companion to the lens, or a companion to the source.
However, as the source is brighter than the clump (see Figure~\ref{fig:allcmd}),
and the error in the free-blending fit is large, we cannot place useful
limits on the lens flux.

The $2.5\,\sigma$ limit $\rho<0.112$, implies $t_*=\rho t_\e<3.1\,$days.
Combined with the measurement $\theta_*=8.5\,\muas$ from Table~\ref{tab:cmd},
this implies $\mu_\rel = \theta_*/t_* > 1.00\,\masyr$.  According to 
Equation~(\ref{eqn:probmu}), this excludes $\sim 0.4\%$ of Galactic events.  
Hence,
while we will include the $\rho$ constraint (or, rather the corresponding
constraint $\theta_\e > 76\,\muas$) in the Bayesian analysis of
Section~\ref{sec:phys-kb180030}, the only real constraint will be from the
measurement of $t_\e$.

We note that {\it Gaia} \citep{gaia16,gaia18}
reports a source proper motion of
$\bmu_S(N,E) = (-7.79,-10.93)\pm (0.45,0.67)\,\masyr$.  If correct,
the source would be moving at about $8.3\,\masyr$ relative to the bulge
proper-motion centroid.  Only of order 1.6\% of bulge sources are moving
this fast.  However, the {\it Gaia} RUWE parameter is 2.2, 
indicating that the measurement
may not be reliable.  Hence, because the measurement is both unusual and
possibly unreliable, we do not include it in the Bayesian analysis.  In any 
case, because there is no significant constraint on $\mu_\rel$, the {\it Gaia}
measurement would not have much effect even if it were included.

\subsection{{KMT-2018-BLG-0087}  
\label{sec:cmd-kb180087}}

The source analysis is overall similar to the case of KMT-2018-BLG-0030
(see Section~\ref{sec:cmd-kb180030}).  
The source position lies within  $\sim 23\,\mas$
of the baseline object (consistent with measurement error), while the
blended light is consistent with zero at $<2\,\sigma$. Hence, if there is
blended light, it would again be associated with the event.  However, again,
because the source is above the clump, while the blend could have at least
10\% of the source light, no useful constraint can be put on lens light.
Because the color and magnitude of the blend are not well-constrained
(and because there is no clear evidence that the blend flux 
is different from zero),
we do not display the blended light in Figure~\ref{fig:allcmd}.

The upper limit $\rho<0.110$ (or $\rho<0.096$) 
is essentially the same as for KMT-2018-BLG-0030,
but in this case, it leads to a much shorter limit, $t_*< 0.50\,$days,
and so to a proper-motion constraint, $\mu_\rel = \theta_*/t_* > 7.0\,\masyr$.  
This is significant and thus can play an important role in the Bayesian
analysis in Section~\ref{sec:phys-kb180087}.  In fact, we will use the
full $\chi^2$ versus $\rho$ envelope, rather than a simple upper limit.
Unfortunately, while {\it Gaia} has an entry for the
source star, it does not report a parallax and proper-motion solution.  
However, this means that future iterations of the {\it Gaia} catalog 
may report a proper-motion measurement.

\subsection{{KMT-2018-BLG-0247} 
\label{sec:cmd-kb180247}}

The source color measurement using KMTC led to a relatively blue color
$(V-I)_{0,S} = 0.51\pm 0.07$.  Hence, we checked the result using
KMTS data, but found a consistent result with a substantially larger error,
$(V-I)_{0,S} = 0.57\pm 0.16$.  The results reported in Table~\ref{tab:cmd}
and shown in Figure~\ref{fig:allcmd} are the weighted average of the
two measurements.  Because of the dearth of such blue stars in the bulge
(even at its location on the turnoff), it is likely that the true
color is 1--1.5$\,\sigma$ redder.  Nevertheless, we consider that our
usual error treatment adequately allows for such variations.

As noted in Section~\ref{sec:anal-kb180247}, the two solutions have
substantially different $\rho$ measurements due to the different source
angles relative to the caustic entrance.  These lead to Einstein-radius
and proper-motion measurements,
\begin{equation}
\theta_\e 
= 0.335\pm 0.045\,\mas;
\qquad \mu_\rel 
= 11.6\pm 1.6\,\masyr
\qquad ({\rm close}),
\label{eqn:thetae_0247_close}
\end{equation}
and
\begin{equation}
\theta_\e 
= 0.256\pm 0.034\,\mas;
\qquad \mu_\rel 
= 8.8\pm 1.2\,\masyr
\qquad ({\rm wide}).
\label{eqn:thetae_0247_wide}
\end{equation}

Figure~\ref{fig:allcmd} shows that the blend is about 0.6 brighter and
0.2 mag bluer than the source.  We find that the baseline object
is separated by about 180 mas from the source, implying that the
blend is about 300 mas from the source.  We can robustly say that
the lens is fainter than this blend, $I_{L,\rm pyDIA}>I_B=20.46$ 
because a lens that
was even within a few tenths of a mag of $I_B$ would ``force''
the additional star that is responsible for the astrometric offset
to be sufficiently separated to be resolved.  However, this limit
has almost no impact on bulge lenses because the field extinction,
$A_I=2.99$, implies that this limit only excludes bulge stars with $M_I\la 2.8$,
which are extremely rare.  Moreover, disk lenses that are sufficiently
massive to provide this light are already heavily disfavored by the
relatively small $\theta_\e$ in Equations~(\ref{eqn:thetae_0247_close})
and (\ref{eqn:thetae_0247_wide}).  Nevertheless, we include this
limit in Section~\ref{sec:phys-kb180247} for completeness.

\subsection{{OGLE-2018-BLG-0298} 
\label{sec:cmd-ob180298}} 

\begin{figure}
\epsscale{0.8}
\plotone{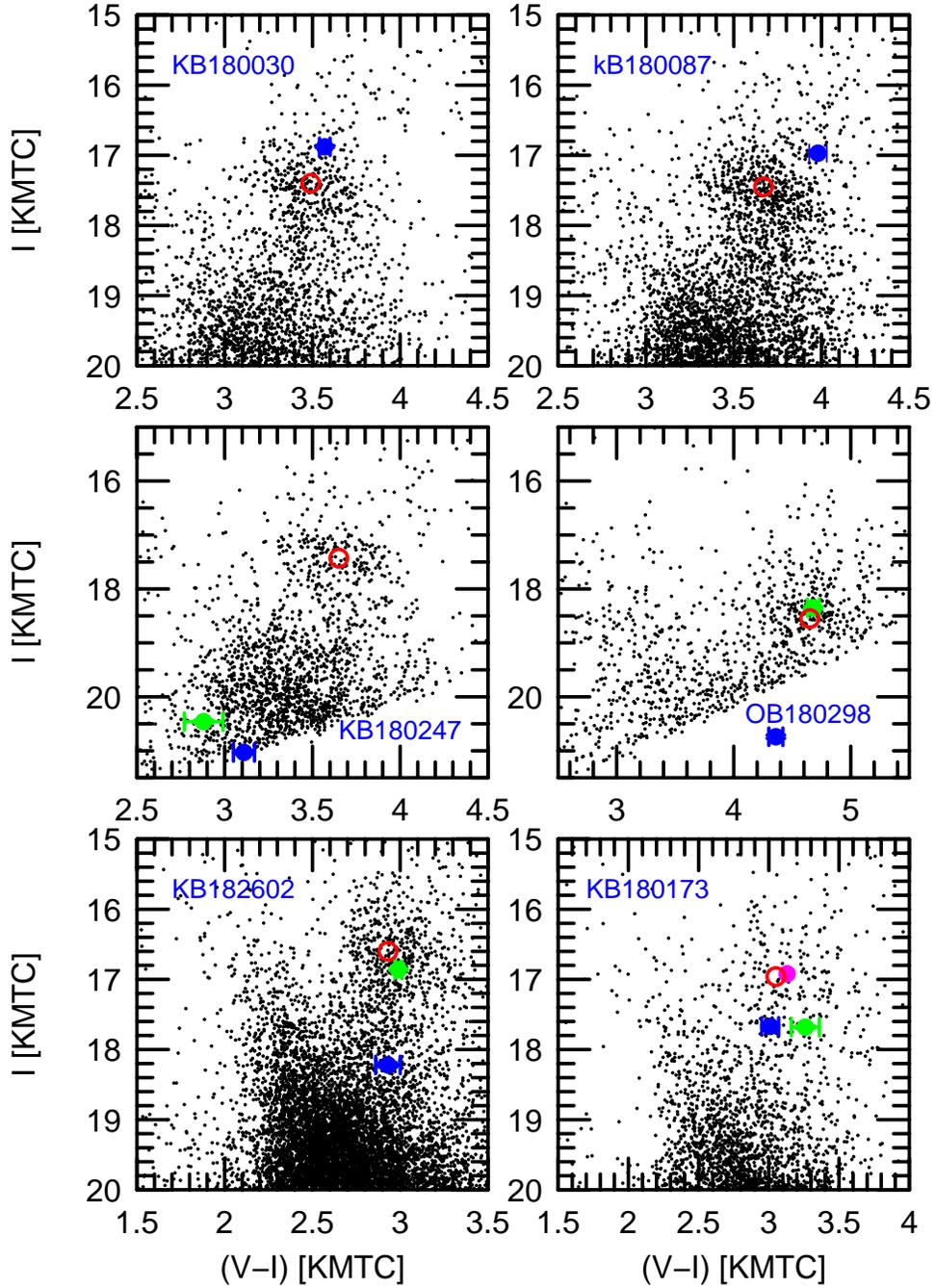}
\caption{Color-magnitude diagrams for 6 of the events analyzed in
this paper, each identified by an abbreviation, e.g.,  KB180030 for
KMT-2018-BLG-0030.  The centroid of the red clump and the lens position
are always shown in red and blue, respectively.  Where relevant, the
blended light is shown in green.  In one case (KB180173), we show
the baseline object in magenta.  When there are multiple solutions,
we show only the source and blend for the lowest-$\chi^2$ solution.
}
\label{fig:allcmd}
\end{figure}

As for all other events in this paper, the OGLE-2018-BLG-0298 CMD
is shown for $[(V-I),I]$.  However, while the clump is clearly visible in 
Figure~\ref{fig:allcmd}, the lower part of the clump 
merges into the background noise, which is due to high extinction.
In principle, this could make it 
difficult to properly measure the mean $I$-band magnitude of the clump.
On the other hand, the clump is approximately horizontal (which is the
expected behavior provided that there is relatively little differential
extinction across the field), implying that the mean $(V-I)$ color of the
clump is independent of height and so not significantly affected by
growing noise at faint magnitudes.

\begin{figure}
\epsscale{0.8}
\plotone{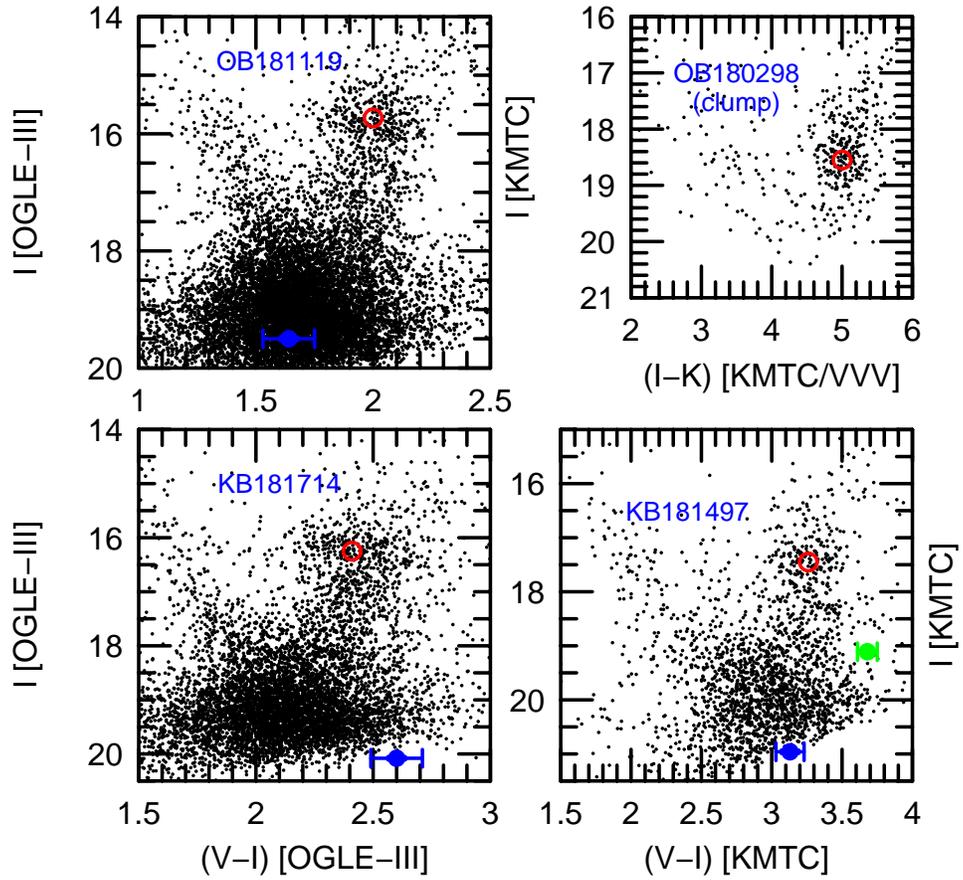}
\caption{Same as Figure~\ref{fig:allcmd}, but for the remaining 3 events
(normal-sized panels).  The undersized panel shows the determination
of $I_{\rm cl}$ from an $[(I-K),I]$ CMD.  See Section~\ref{sec:cmd-ob180298}.
}
\label{fig:allcmd2}
\end{figure}

We therefore measure the height, $I_{\rm cl}$, (but not the color) of the
clump from an $[(I-K),I]$ diagram, which we construct by matching the
KMTC pyDIA $I$-band measurements to the $K$-band measurements from the
VVV catalog \citep{vvv-survey1,vvvcat}.  This measurement is illustrated
in the ``undersized panel'' of Figure~\ref{fig:allcmd2}.  We emphasize
that the only purpose of this panel is to measure $I_{\rm cl}$, which
is then reproduced in Figure~\ref{fig:allcmd} and in Table~\ref{tab:cmd}.

As shown in Figure~\ref{fig:allcmd}, the source is blended with
a clump giant, compared to which it is about 2.5 mag fainter.
The clump giant is separated from the source by about 400 mas,
so it is unlikely to be associated with the event.  However, it
does prevent us from placing any useful limits on the lens light.

From Table~\ref{tab:ob0298parms}, it would appear that $\rho$ is measured with
tolerable precision, albeit with substantially different error bars
for the two solutions.  However, the $\chi^2(\rho)$ function is
highly non-Gaussian (i.e., not even approximately quadratic).
Rather, for, e.g., the close solution, it is approximately
flat for $0.003\la \rho \la 0.0045$, rising linearly,
$\Delta\chi^2\simeq 12 - 4000\rho$ at lower values, and rising
almost vertically at higher values.  If we simply adopt the best-fit
$\rho\sim 3.6\times 10^{-3}$, then $\theta_\e\simeq 0.42\,\mas$ and
$\mu_\rel \simeq 4.8\,\masyr$.  However, because somewhat lower values of $\rho$
(hence, higher $\mu_\rel$, which are kinematically favored by
Galactic models) have a minimal $\chi^2$ penalty, we will use the 
actual $\chi^2(\rho)$ function in the Bayesian analysis of 
Section~\ref{sec:phys-ob180298}, rather than the Gaussian approximation.

\subsection{{KMT-2018-BLG-2602} 
\label{sec:cmd-kb182602}}

Very similarly to the case of OGLE-2018-BLG-0298 
(Section~\ref{sec:cmd-ob180298}), the source is blended with a clump
giant, although in this case the source is 1.5 mag fainter.  In addition,
in this case, the source is aligned with the baseline object to within
$10\,\mas$, probably indicating that the (clump-giant) blend is a companion
to the (sub-giant) source.  If one were to naively apply the arguments
given above for KMT-2018-BLG-0030 and KMT-2018-BLG-0087, one might conclude
from the presence of a clump-giant blend that no useful limits could
be put on lens light.  However, this proves not to be the
case (see Section~\ref{sec:phys-kb182602}).

In Table~\ref{tab:kb2602parms}, 
we list a $2.5\,\sigma$ limit $\rho<0.074$, which considered
by itself would imply $\mu_\rel>0.13\,\masyr$, i.e., completely unconstraining
according to Equation~(\ref{eqn:probmu}).  We check whether there are
are additional subtle structures in the $\chi^2(\rho)$ function that might
provide additional constraints.  However, we find, on the contrary,
that all values $0<\rho< 0.05$ are consistent with the minimum at $1\,\sigma$.
Hence, we simply adopt the Table~\ref{tab:kb2602parms} 
$2.5\,\sigma$ limit in the Bayesian analysis in Section~\ref{sec:phys-kb182602}.

For completeness, we note that {\it Gaia} lists a proper-motion measurement
for the baseline object of
$\bmu_{\rm base}(N,E) = (-9.89\pm 0.12,-2.37\pm 0.19)\,\masyr$.
If, as appears very likely, the baseline object is composed of the
source and its companion, then the source proper motion is
given by $\bmu_S = \bmu_{\rm base}$.  This source proper motion would
then be about $|\Delta\bmu|\sim 4.7\,\masyr$, i.e., $1.6\sigma$ from
the bulge mean.  However, we do not include this measurement in the
analysis for two reasons.  First, we cannot be certain that the blend is
a companion to the source.  Second, in the absence of significant
constraints on $\mu_\rel$, knowledge of the source proper motion plays
very little role.

\subsection{{OGLE-2018-BLG-1119} 
\label{sec:cmd-ob181119}} 

OGLE-2018-BLG-1119 is one of only two events that are analyzed in
this paper that are covered by OGLE-III.  We therefore provide calibrated
photometry for it in Figure~\ref{fig:allcmd2} and in Table~\ref{tab:cmd}.

Moreover, this also allows us to compare the field-star astrometry
of pyDIA KMTC to that of OGLE-III.  Both catalogs find a neighbor roughly
due east of the source, at about 760 mas for OGLE-III and 870 mas for KMTC.
However, the two catalogs divide the light between these two stars
differently:  OGLE-III puts 75\% of the total light in the source,
whereas for KMTC, this figure is 33\%.  Furthermore, the combined light
of the two stars is 0.32 mag brighter in KMTC (after alignment of the
two systems).

From this comparison, we can robustly conclude that the baseline object 
is semi-resolved from its neighbor, but we cannot tell, a priori,
whether OGLE-III or KMTC has a more accurate assessment of the brightness 
of the baseline object.  Nevertheless, it is striking that the KMTC 
baseline-object measurement is nearly identical to the source flux as 
determined from the light-curve model, implying that it is quite possible
that there is very little blended light, e.g., from the lens.  Therefore,
we do not show any blend in the CMD.

On the other hand, we cannot rule out that the OGLE-III measurement is
actually correct.  Therefore, we must place a conservative limit on lens
light, namely that $I_L > 19.7$, i.e., roughly as bright as the source.

In Table~\ref{tab:ob1119parms}, 
we show only an upper limit for the $\rho$ measurement.
This is because, while there is a clearly defined minimum in $\chi^2(\rho)$
at the $1\,\sigma$ level, all values $\rho<0.065$ are consistent at 
$\Delta\chi^2<4$.  If we, for the moment, take the best fit value 
$\rho=0.045$ seriously, then $t_*= 1.84\,$days, so that 
$\mu_\rel= \theta_*/t_*=0.14\,\masyr$, which is extraordinarily unlikely
according to Equation~(\ref{eqn:probmu}).  Therefore, the apparent 
1-$\sigma$ ``measurement'' of $\rho$ is almost certainly due to a
statistical fluctuation or minor systematics.  Hence, in 
Section~\ref{sec:phys-ob181119}, we use only the upper limit $\rho<0.071$
(or $\rho<0.061$).

\subsection{{KMT-2018-BLG-0173} 
\label{sec:cmd-kb180173}}

Recall from Section~\ref{sec:anal-kb180173} that it was difficult
to distinguish among three possibilities: 2L1S (close), 2L1S (wide), and 1L2S.
In brief, 2L1S (close) was preferred over the other two by 
$\Delta\chi^2\sim 11$.  Moreover, both the 2L1S (wide) and 1L2S solutions
had reasonably well-determined $\rho$ values that would imply improbably
low values of $\mu_\rel$.  Note that the best fit value for $\theta_*$ given
in Table~\ref{tab:cmd} is (as always) for the lowest-$\chi^2$ solution.
Hence, $\theta_{*,\rm wide} = \theta_{*,\rm close}\times 10^{0.48/5} = 5.45\,\muas$.

Nevertheless, we found that these proper-motion arguments could be evaded
in both cases by accepting a modest $\chi^2$ penalty for enforcing $\rho=0$,
which would thereby accommodating much higher (including very plausible) proper
motions.

An additional color-based argument was given against 1L2S, but unfortunately
this was statistically weak; $\Delta\chi^2=2.5$.

Here, we show that the CMD, when combined with the astrometric measurements, 
argues in exactly the opposite direction, i.e., that both the 2L1S (wide) and 
1L2S solutions are strongly preferred over the 2L1S (close) solution.

The argument concerns the blended light.  The first point is that in
the 2L1S (close) solution, the blended light is about equal to that
of the source, while in both the 2L1S (wide) and 1L2S solutions, the
blended light is consistent with zero at $1\,\sigma$.  This difference
becomes important because the measured offset between the source and
the baseline object is just 7 mas, i.e., within the measurement error.
This close alignment implies that the blended light (if any) is
almost certainly associated with the lens, either the lens itself,
a companion to the lens, or a companion to the source.  (The usual
argument against field-star contamination is made even stronger by
the fact that the surface density of clump stars (which are proxies for the
giant stars of direct interest here) is only half that of Baade's
Window, and so a factor 3--5 smaller than that of typical 
events\footnote{We evaluate this using the measurements by
\cite{nataf13} in the neighborhood of the reflection point
$(l,b)\rightarrow (l,-b)$ because the northern bulge is not well covered
in that study.}.)

Figure~\ref{fig:allcmd} shows the source and blend positions for the
2L1S close solution in blue and green, respectively.  The magenta
point shows the baseline object, which would be the source position for
either of the other solutions, assuming no blending.  Thus, in the 
2L1S close solution (but for neither of the others), the blend would
be (within measurement error) a twin of the source, and thus almost
certainly a companion to the source of virtually identical mass
in a very rapid phase of stellar evolution.

Because each of these three solutions has some very improbable feature,
we do not regard it as possible to confidently choose among them
based on current evidence.  It is possible that additional data, such
as AO imaging and/or spectroscopy on large telescopes, will eventually
resolve the issue.  However, at present, we believe that this event
should not be cataloged as ``planetary''.

\subsection{{KMT-2018-BLG-1497} 
\label{sec:cmd-kb181497}}

In Section~\ref{sec:anal-kb181497}, we showed that there are three
different planetary solutions for KMT-2018-BLG-1497,
with mass ratios, $q$, spanning two orders
of magnitude, plus a 1L2S solution, all within $\Delta\chi^2\la 4$.
Thus, the $\theta_*$ evaluation in Table~\ref{tab:cmd} and the CMD in
Figure~\ref{fig:allcmd2} (both, as usual, for the lowest-$\chi^2$ solution),
are shown only for completeness: they could become of interest if
future observations can distinguish among these solutions.

\subsection{{KMT-2018-BLG-1714} 
\label{sec:cmd-kb181714}}

In Section~\ref{sec:anal-kb181714}, we showed that for KMT-2018-BLG-1714,
there is a 1L1S
solution within $\Delta\chi^2<1$ of the best planetary solution.
In addition, there was a factor 3.5 degeneracy in $q$ at $\Delta\chi^2\sim7$.
Thus,  as for KMT-2018-BLG-1497,
the $\theta_*$ evaluation in Table~\ref{tab:cmd} and the CMD in
Figure~\ref{fig:allcmd2} are shown only for completeness because
they could become of interest if
future observations can distinguish among these solutions.

\section{{Physical Parameters}
\label{sec:phys}}

To make Bayesian estimates of the lens properties, we follow the same
procedures as described in Section~5 of \citet{2018prime}.  We refer the
reader to that work for details.

In Table~\ref{tab:physall}, we present the resulting Bayesian estimates
of the host mass $M_{\rm host}$, the planet mass $M_{\rm planet}$, 
the distance to the lens system $D_L$, and the planet-host projected
separation $a_\perp$.  For the majority
of events, there are two or more competing solutions.  For these cases
(following \citealt{2018prime}),
we show the results of the Bayesian analysis for each solution separately,
and we then show the ``adopted'' values below these.  For $M_{\rm host}$,
$M_{\rm planet}$, and $D_L$, these are simply the weighted averages of the
separate solutions, where the weights are the product of the two
factors at the right side of each row.  The first factor is simply
the total weight from the Bayesian analysis.  The second is 
$\exp(-\Delta\chi^2/2)$ where $\Delta\chi^2$ is the $\chi^2$ difference
relative to the best solution.  For $a_\perp$,
we follow a similar approach provided that either the individual solutions
are strongly overlapping or that one solution is strongly dominant.  
If neither condition were met, we would enter ``bi-modal'' instead.
However, in practice, this condition is met for all 4 events for which
there is potentially an issue.

\begin{figure}
\epsscale{1.0}
\plotone{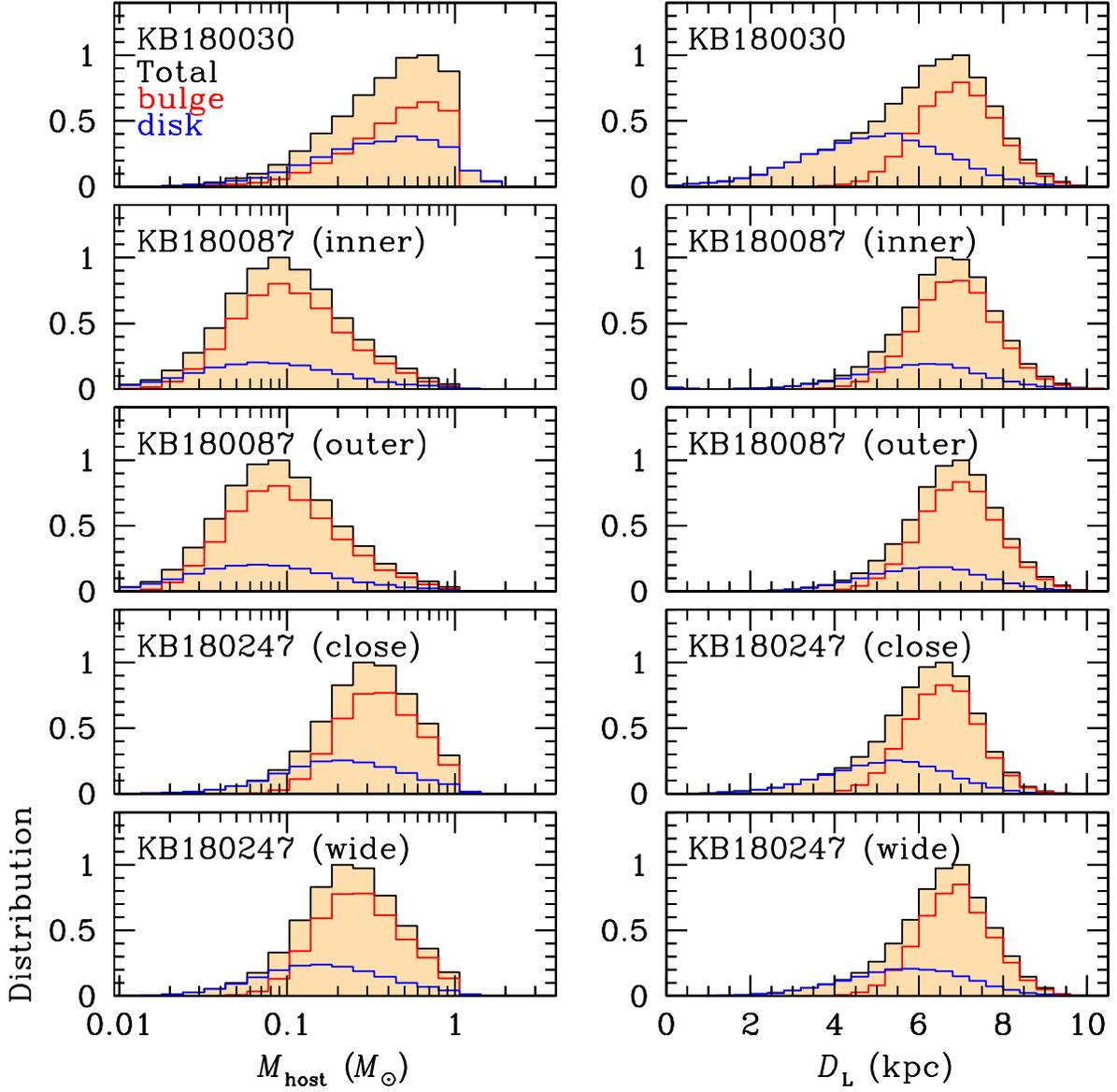}
\caption{Histograms of the host mass (left) and lens distance (right)
for 3 of the 6 unambiguously planetary events, as derived from the 
Bayesian analysis.  Disk (blue) and bulge (red) distributions are shown
separately, with their total shown in black.
}
\label{fig:bayes1}
\end{figure}

We present Bayesian analyses for 6 of the 9 events, but not for
KMT-2018-BLG-0173, KMT-2018-BLG-1497, and KMT-2018-BLG-1714, for which
we cannot distinguish between competing interpretations of the event.
See Sections~\ref{sec:anal-kb180173}, \ref{sec:anal-kb181497}, and
\ref{sec:anal-kb181714}.
Figures~\ref{fig:bayes1} and \ref{fig:bayes2} show histograms
for $M_{\rm host}$ and $D_L$ for these 6 events.

\subsection{{KMT-2018-BLG-0030}
\label{sec:phys-kb180030}}

For KMT-2018-BLG-0030, there is only one light-curve solution.  As discussed
in Sections~\ref{sec:anal-kb180030} and \ref{sec:cmd-kb180030}, while we
include two constraints, $t_\e = 27.94\pm 0.11\,$ days and
$\theta_\e> 76\,\muas$, only the first of these has practical importance.

As illustrated by Figure~\ref{fig:bayes1}, the relatively long timescale 
somewhat favors more massive hosts, but the lens-distance distribution
primarily just reflects the Galactic prior.

\subsection{{KMT-2018-BLG-0087}
\label{sec:phys-kb180087}}

For KMT-2018-BLG-0087, there are two solutions.  For each, there are
two constraints, one on $t_\e$ (as given in Table~\ref{tab:kb0087parms})
and the other on $\theta_\e$.  As discussed in Section~\ref{sec:anal-kb180087},
the seemingly crude limit on $\rho$ is significant due to the
short $t_\e$.  Therefore, as discussed in Section~\ref{sec:anal-kb180087},
the profile $\chi^2(\rho)$ actually matters.  Therefore, we implement
the $\theta_\e$ constraint by, for each simulated event with
Einstein radius $\theta_\e$, first evaluating $\rho=\theta_*/\theta_\e$
(where $\theta_* = 9.53\,\muas$), and then assigning a weight
\begin{equation}
w_{\theta_\e} = {\exp[-\chi^2(\rho)/2]\over Q};
\qquad Q\equiv \int_0^\infty d\rho\exp[-\chi^2(\rho)/2],
\label{eqn:thetae-kb180087}
\end{equation}
where
$$
\chi^2_{\rm inner}(\rho) = \biggl({\rho\over 0.044}\biggr)^2;
$$
\begin{equation}
\chi^2_{\rm outer}(\rho) = \biggl({\rho\over 0.080}\biggr)^2 \quad (\rho<0.09);
\qquad
\chi^2_{\rm outer}(\rho) = {81\over 64} + 
5\biggl({\rho-0.09\over 0.02}\biggr)^2 \quad (\rho>0.09),
\label{eqn:chi2-kb180087}
\end{equation}
with $Q_{\rm inner} = 0.044\sqrt{\pi/2} = 0.055$, and
$Q_{\rm outer} = 0.081$.

The host mass is peaked near the star/brown-dwarf boundary, while
the lens distance is strongly dominated by the bulge population.
See Figure~\ref{fig:bayes1}.
The physical reason for this is that the combination of the
modeling constraints and the Galactic priors strongly favors
events with small $\theta_\e=\sqrt{\kappa M \pi_\rel}$ (so, low $M$ and
low $\pi_\rel$).  That is, by itself, the $2.5\,\sigma$ limit on $\rho$
would only imply a lower limit: $\theta_\e >0.086\,\mas$, but substantially
larger $\theta_\e$ would lead to high proper motions that are heavily
disfavored by the Galactic priors.

\subsection{{KMT-2018-BLG-0247}
\label{sec:phys-kb180247}}

For KMT-2018-BLG-0247, there are two solutions.  For each, there are
three constraints: the first on $t_\e$ (as given in 
Table~\ref{tab:kb0247parms}), the second on
$\theta_\e$ (as given in Equations~(\ref{eqn:thetae_0247_close})
and (\ref{eqn:thetae_0247_wide})), and the third on lens light
$I_{L,\rm pyDIA}>20.46$ (as described in Section~\ref{sec:cmd-kb180247}).
In order to put this limit on a calibrated scale, we
estimate $I_{\rm calib} - I_{\rm pyDIA} = I_{\rm cl,calib} -I_{\rm cl,pyDIA}
= I_{\rm cl,0} +A_I -I_{\rm pyDIA} = -0.03$, where 
$I_{\rm cl,0}$ and $I_{\rm pyDIA}$ are from Table~\ref{tab:cmd} and $A_I=2.99$ 
is from the KMT webpage as described in Section~\ref{sec:anal-kb180030}.
Therefore, the calibrated limit is $I_L> 20.43$.

The host distributions (Figure~\ref{fig:bayes1})
are dominated by bulge M dwarfs.  
\citet{kb190371} showed that for measured proper motions 
$\mu_\rel\la 10\,\masyr$, Bayesian mass estimates depend almost
entirely on the $\theta_\e$ measurement.  Our results for both the
close and wide solutions (which differ by a factor 1.3) are in good
accord with the predictions from their Figures 6 and 7.  Note that,
as anticipated in Section~\ref{sec:cmd-kb180247}, the lens flux constraint
plays almost no role because the other constraints already favor very
faint lenses.

\subsection{{OGLE-2018-BLG-0298}
\label{sec:phys-ob180298}} 


For OGLE-2018-BLG-0298, there are two solutions.  For each, there are
three constraints.  The first, on $t_\e$, is given in 
Table~\ref{tab:ob0298parms}, while each of the other two, on
$\theta_\e$ and $\bpi_\e$, require additional discussion.

As mentioned in Section~\ref{sec:cmd-ob180298}, for the close solution,
$\chi^2(\rho)$ is both fairly broad
and highly non-Gaussian.  Therefore, as in the case of KMT-2018-BLG-0087
(Section~\ref{sec:phys-kb180087})
we employ Equation~(\ref{eqn:thetae-kb180087}) by
directly characterizing $\chi^2(\rho)$
$$
\chi^2_{\rm close}(\rho) = 12-4000\rho \quad (\rho<0.003);
\quad =0\quad (0.003<\rho<0.0045); 
\quad =24000\rho-108\quad (\rho<0.0045),
$$
\begin{equation}
\chi^2_{\rm wide}(\rho) = \biggl({1000\rho-3.98\over 0.48}\biggr)^2;
\label{eqn:chi2-ob180298}
\end{equation}
with $Q_{\rm close} = 2.09\times 10^{-3}$, and
$Q_{\rm wide} = \sqrt{2\pi} 0.48\times 10^{-3}=1.20\times 10^{-3}$.

To incorporate the parallax measurement, we could in principle double the
number of solutions from 2 to 4, and then average together the Bayesian
results from the two close solutions and from the two wide solutions.
However, inspection of the parallax solutions given in 
Section~\ref{sec:anal-ob180298} (in principal-axis format), shows that 
the pair of close (or wide) solutions
differ by much less than their errors, and the error ellipses are also
extremely similar.  These are natural consequences of the fact that
$u_0\ll 1$.  Hence, it make more sense to average these solutions,
before applying the Bayesian results than after.  Next, we notice that
the parallax solutions for the close and wide solutions for a given
sign of $u_0$ are even more similar than are, e.g., the two close
solutions.  Hence, for simplicity, we simply average together the four
solutions (in equatorial coordinates) to obtain
\begin{equation}
a_{i,0} = \left(\matrix{\pi_{\e,N,0} \cr \pi_{\e,E,0}}\right)= 
\left(\matrix{-0.1433 \cr +0.0079}\right);
\qquad
c_{ij} = \left(\matrix{0.4287 & 0.0127\cr 0.0127 & 0.0128 }\right),
\label{eqn:ob180298par}
\end{equation}
and then derive the parallax weight,
\begin{equation}
w_{\rm par} = {\exp(-\sum_i\sum_j (a_i-a_{i,0}) b_{ij}(a_j- a_{j,0})/2)\over 
\sqrt{2\pi{\rm det(b)}}},
\label{eqn:ob180298par2}
\end{equation}
where $b\equiv c^{-1}$ and
$a_i=(\pi_{\e,N},\pi_{\e,E})_i$ of simulated event, $i$.

The results of the Bayesian analysis can be qualitatively understood
by considering a ``typical'' value of $\rho\sim 4\times 10^{-3}$, hence
$\theta_\e\sim 0.38\,\mas$ and $\mu_\rel\sim 4\,\masyr$.  If there
were no parallax information, then the argument of \citet{kb190371} would
lead to a mass estimate $M\sim 0.44 M_\odot$, with most lenses in the
bulge.  At typical bulge $\pi_\rel \sim 16\,\muas$, lenses at this
peak mass would have $\pi_\e\sim 0.07$, which is quite compatible
with the parallax constraints.  However, at a factor 2 below this
peak, the parallax constraint $|\pi_{\e,\parallel}|\la 0.11$ starts to
suppress many simulated events.  Hence, the mass peak in 
Figure~\ref{fig:bayes2} is shifted higher relative to the \citet{kb190371}
prediction.

\subsection{{KMT-2018-BLG-2602}
\label{sec:phys-kb182602}}

\begin{figure}
\epsscale{1.0}
\plotone{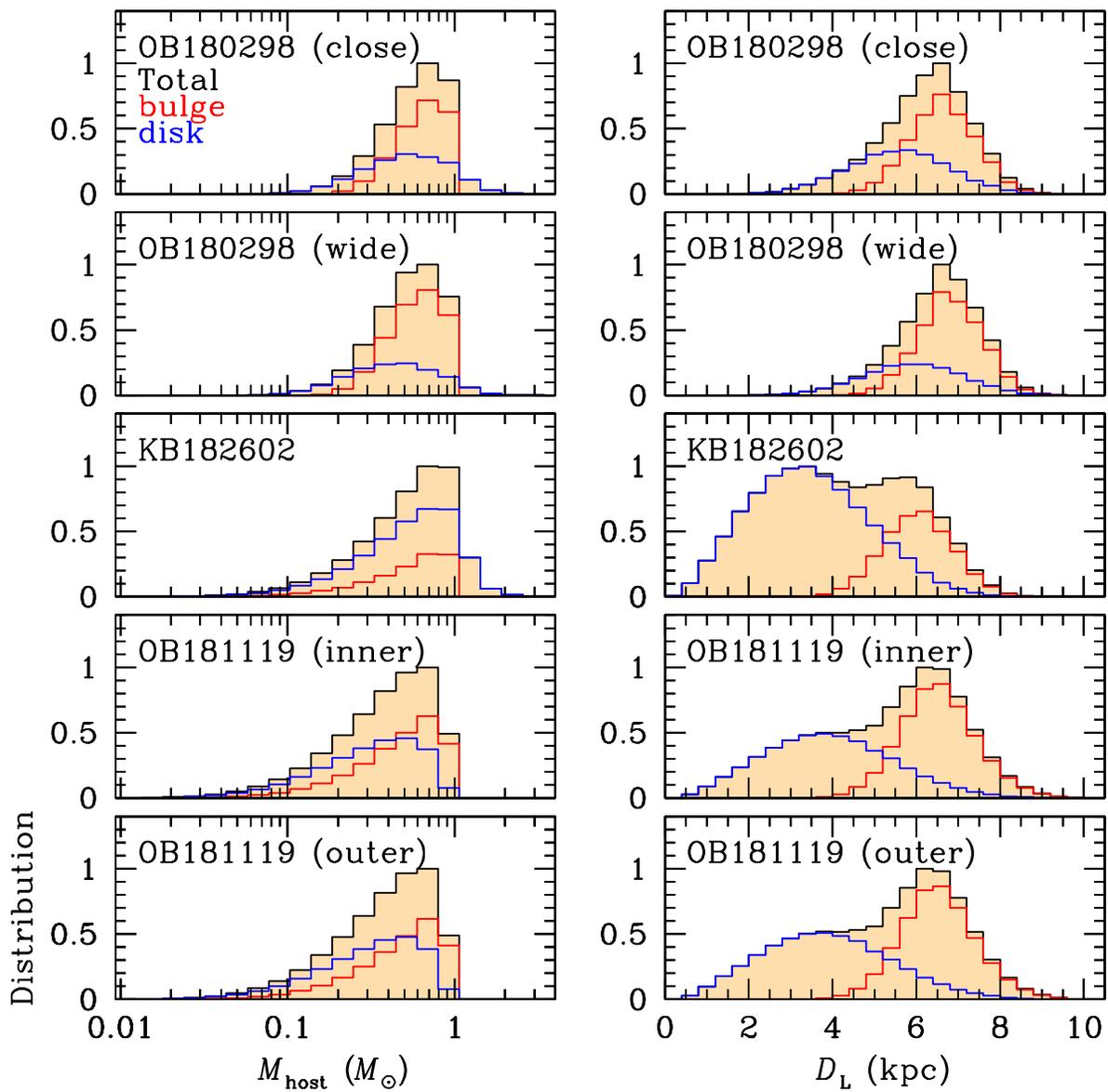}
\caption{Same as Figure~\ref{fig:bayes1}, except that this figure shows
the remaining 3 unambiguously planetary events.
}
\label{fig:bayes2}
\end{figure}

As discussed in Section~\ref{sec:anal-kb182602}, we consider that
its $\Delta\chi^2=10.3$ preference decisively resolves the degeneracy
in favor of the outer solution.  When we initially carried out the Bayesian
analysis, we considered that there were only two
constraints, one on $t_\e=99\pm 14\,$day 
(given by Table~\ref{tab:kb2602parms}) and the
other (from Tables~\ref{tab:kb2602parms} and \ref{tab:cmd}), 
$\theta_\e > 0.043\,\mas$.  Moreover, as discussed in 
Section~\ref{sec:anal-kb182602}, the latter has almost no impact and
is included only for completeness.  In particular, we did not consider 
as relevant the fact that the lens must be fainter than the clump-giant blend.
However, the resulting lens-distance histogram,
which was qualitatively similar to the finally adopted one that
is shown in Figure~\ref{fig:bayes2}, contained a substantial fraction
of nearby lenses that, at least potentially, might be excluded by
even such a weak constraint.  We therefore calibrated the pyDIA
measured blend flux $[(V-I),I]_B=(2.99,16.86)$ to obtain
$I_L>16.82$ and $V_L>19.57$, using the same procedures as in 
Section~\ref{sec:phys-kb180247}.  For this purpose, we used $A_I=2.32$
from the KMT webpage and $E(V-I) = A_I/1.10$, estimated from Figure~6
of \citet{nataf13}.  We find that these constraints have a small,
but non-negligible effect, as we quantify below.

The bi-modal distance distribution in Figure~\ref{fig:bayes2}, can be
understood as follows.  For typical $\mu_\rel \sim 6\,\masyr$,
$\sqrt{\kappa M \pi_\rel}=\theta_\e = \mu_\rel t_\e\sim 1.7\,\mas$, which implies
massive, nearby lenses, e.g., $M=0.7\,M_\odot$, $D_L=1.6\,\kpc$.  Hence,
a large number of simulated disk events are consistent with this timescale.
On the other hand proper motions that are, e.g., 5 times slower,
so $\theta_\e\sim 0.33\,\,\mas$ would be consistent with massive lenses
in the bulge.  Such low proper motions are somewhat disfavored, but this
is compensated by the high surface density of bulge stars compared to
disk stars.  Note, however, that in both cases, relatively massive
lenses are favored, as shown in the left panel.

As noted above, before including the lens-flux constraints, the Bayesian
priors would favor large $\theta_\e$ and hence massive, nearby lenses.
Specifically, we find that this suppression of nearby disk lenses reduces
the median value of $M_{\rm host}$
by $10\%$ relative to the case of no flux constraint, while the median 
value of $D_L$ is increased by $8\%$.

\subsection{{OGLE-2018-BLG-1119}
\label{sec:phys-ob181119}} 

For OGLE-2018-BLG-1119, there are two solutions.  For each, there are
three constraints: the first on $t_\e$ (as given in 
Table~\ref{tab:ob1119parms}), the second on
$\theta_\e$ (as derived from Tables~\ref{tab:ob1119parms} and
\ref{tab:cmd}), i.e., $\theta_{\e,\rm inner} > 0.062\,\mas$ and
$\theta_{\e,\rm outer} > 0.063\,\mas$,
and the third on lens light
$I_L>19.7$ (as described in Section~\ref{sec:cmd-ob181119}).
We note that this limit is already on the OGLE-III calibrated scale.

As in the case of KMT-2018-BLG-2602, the distance distribution is
bi-modal, and the explanation for this is similar: the event has a
relatively long $t_\e \sim 40\,$days and there is only an extremely
weak constraint on $\theta_\e$.  However, following the logic of 
Section~\ref{sec:phys-kb182602} and noting that the timescale is
significantly shorter than in that case, we expect the disk peak to
be weaker and the bulge peak to be stronger, while we expect the
median mass to be significantly lower.  All these expectations are
confirmed by Figure~\ref{fig:bayes2} and Table~\ref{tab:physall}.

.\section{{Discussion}
\label{sec:discussion}}

\subsection{{Summary of 2018 AnomalyFinder Detections}
\label{sec:summary}}

The main goal of this paper has been to complete the 2018 AnomalyFinder
sample, with the ultimate purpose being to lay the basis for
a mass-ratio function analysis.  We summarize this work in 
Table~\ref{tab:all2018events}, where we combine the results on the
9 planetary (or possibly planetary) events analyzed here with 10
previously published such events that were identified by the 
AnomalyFinder algorithm.  We have divided these into 14 events that
are likely to survive final selection for mass-ratio studies and 5 that
are unlikely to survive.  However, in this paper we do not give final
designations for individual events but, rather, provide the necessary
information for others to do so.  In Table~\ref{tab:all2018events},
we indicate when there are multiple solutions and give the $(\log q,s)$
of the lowest $\chi^2$ solution.  In future mass-ratio studies,
it will be necessary to define a best-representative $\log q$ and to exclude
events for which the different solutions are too divergent to do so.

We note that one previously published planetary event, 
OGLE-2018-BLG-1996, was identified by the AnomalyFinder algorithm but 
was not selected in the by-eye review.  This is one of only two
such cases out of all previously published planets from 2016-2019
that were recovered by the AnomalyFinder algorithm.  
Because the number of such failures is small, they do not
generate a significant systematic effect.  We therefore believe
that this event should be included in the mass-ratio function sample.
Again, however, it is not the purpose of this paper to make a final
decision on this issue.

Table~\ref{tab:all2018events} should be compared to Table~14 of 
\citet{2018prime}, which contains a total of 26 planetary (or possibly
planetary) events.  Three of those events (below the double line)
are clearly unsuitable for mass-ratio studies.  In addition, as
noted by \citet{2018prime}, the OGLE-2018-BLG-1700 planet 
was discovered in a binary system, which makes it subject to different
selection biases.  Further, the 1L2S/2L1S degeneracy of OGLE-2018-BLG-1544
are likely to be too severe to include it, while the mass-ratio uncertainties
of OGLE-2018-BLG-1025 and OGLE-2018-BLG-1126 are too severe to include them.  
Hence, it is plausible that a total of approximately 33 (14 sub-prime-field
and 19 prime-field planets) will
be available from 2018 alone for mass-ratio studies.  This would
be the largest microlensing planet sample to date.

\subsection{{6-D Distribution}
\label{sec:6D}}
\begin{figure}
\epsscale{1.0}
\plotone{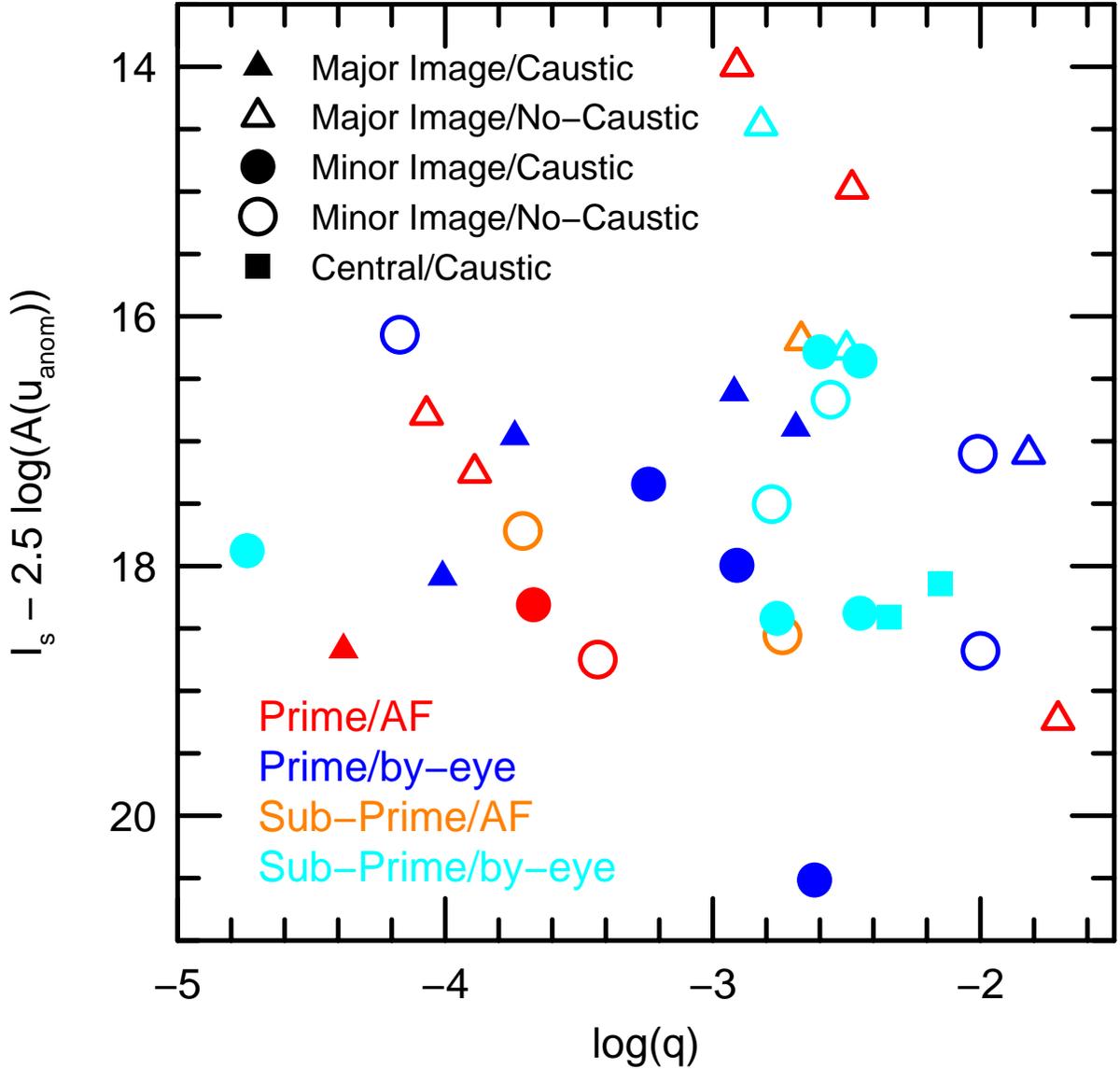}
\caption{6-dimensional scatter plot of 33 planetary events.
(1) Abscissa: log mass ratio.
(2) Ordinate: source magnitude of unperturbed event at time of anomaly.
(3) Primary (red, blue) vs.\ non-primary (orange, cyan) colors: 
prime vs.\ sub-prime fields.
(4) Reddish (red, orange) vs.\ bluish (blue, cyan) colors:
AnomalyFinder vs.\ by-eye discoveries.
(5) Filled vs.\ open symbols: caustic-crossing vs.\ non-caustic-crossing 
anomalies.
(6) Shape: Major-image (triangles), minor-image (circles), and
central-caustic (squares) perturbations.  The two most important patterns
are: (A) a threshold of detections at $I_S - 2.5\log[A(u_{\rm anom})]=18.75$
and (B) the dearth of by-eye discoveries of non-caustic-crossing anomalies
(bluish open symbols) for $\log q<-3$.
}
\label{fig:6d}
\end{figure}

In Figure~\ref{fig:6d}, we show a six-dimensional (6-D) representation of these
33 planets, including 2 continuous dimensions (given by the axes)
and 4 discrete dimensions that are represented by colors and point types.
The abscissa and ordinate are $\log q$ and 
$I_{S,\rm anom}\equiv I_S - 2.5\log[A(u_{\rm anom})]$,
with the latter being the source brightness in the unperturbed event
at the time of the anomaly.  Planets from the prime fields are marked
in primary colors (red and blue), while planets from the sub-prime fields 
are marked in non-primary colors (orange and cyan).
Planets that were discovered by AnomalyFinder are marked in reddish colors
(red and orange), while those previously identified by eye
are marked in bluish colors (blue and cyan).  Planets with major-image
anomalies are shown as triangles, those with minor-image
anomalies are shown as circles, while the two that cannot be classified
as either (KMT-2018-BLG-0247 and OGLE-2018-BLG-0740)
are shown as squares.  Events for which the source crossed a
caustic are shown as filled symbols, while those for which it did not
are shown as open symbols.

The most striking feature of this diagram is the apparent threshold
of AnomalyFinder detection at $I_{S,\rm anom}= 18.75$, with one major
exception (OGLE-2018-BLG-0962) and one minor exception (KMT-2018-BLG-2718),
both being relatively high-$q$ planets.
Another very striking feature is the paucity of by-eye detections
of non-caustic-crossing events (open bluish symbols) at low-$q$:
i.e., 1 (OGLE-2018-BLG-1185) out of 5 for $\log q<-3$ compared to
7 out of 12 for $\log q>-3$.  
It is also notable that among the 16 caustic-crossing events, all but
two were discovered by eye.  Moreover, both of the AnomalyFinder
discoveries (OGLE-2018-BLG-0383 and OGLE-2018-BLG-0977) were in
prime fields and at $\log q <-3$, a regime where machines may do better
than people because the relatively weak signals of low-$q$ events
are spread out over a greater number of data points.
That is, it appears that AnomalyFinder
was essential to finding low-$q$ events, both with and without caustic
crossings.

\citet{zhu14} predicted that roughly half of all planet detections
in a KMT-like survey would not have caustic-crossing features.
The 2018 AnomalyFinder sample, which has 16 caustic-crossing and 17 
non-caustic-crossing events, is consistent with this prediction.

Another anticipated feature of the KMT survey is
also confirmed.  When KMT began regular observations in 2016, it
adopted the layered approach pioneered by OGLE, in which a relatively
small region would be monitored at high-cadence (which we call the 
``prime fields'') and much larger regions would be monitored at a series
of lower cadences \citep{eventfinder}.  
It was expected that the higher-cadence fields would be more sensitive 
to lower-$q$ planets \citep{henderson14}.
Figure~\ref{fig:6d} shows that for $\log q<-3$,
9 of 11 planets are from prime fields (primary colors), compared
to 10 out of 22 for $\log q>-3$.

Finally, there are 14 planets with major-image perturbations, compared
to 17 with minor-image perturbations, which is statistically
consistent with the expectation that these should be about equal.

\acknowledgments 
This research has made use of the KMTNet system operated by the Korea
Astronomy and Space Science Institute (KASI) and the data were obtained at
three host sites of CTIO in Chile, SAAO in South Africa, and SSO in
Australia.
This research was supported by the Korea Astronomy and Space Science Institute 
under the R\&D program(Project No. 2022-1-830-04) supervised by the Ministry of Science and ICT. 
Work by C.H. was supported by the grants of National Research Foundation 
of Korea (2020R1A4A2002885 and 2019R1A2C2085965).
J.C.Y. acknowledges support from US NSF Grant No. AST-2108414.
W.Z. and H.Y. acknowledge support by the National Science Foundation of China (Grant No. 12133005).

 \begin{deluxetable}{lrrrrrr}
 \tablecolumns{7} \tablewidth{0pc}
 \tablecaption{\textsc{Event Names, Cadences, Alerts, and Locations}}
 \tablehead{\colhead{Name} & 
\colhead{$\Gamma\,({\rm hr}^{-1})$} &
\colhead{Alert Date} &
\colhead{RA$_{\rm J2000}$} &
\colhead{Dec$_{\rm J2000}$} &
\colhead{$l$} &
\colhead{$b$} }
 \startdata
KMT-2018-BLG-0030 & 1.0 & 21 Jun 2018 & 17:38:04.00 & $-28$:02:29.85  & $-0.11$  & $+1.88$  \\
\hline
KMT-2018-BLG-0087&  2.0 & 21 Jun 2018 & 17:37:18.48 & $-27$:49:55.42  & $-0.03$  & $+2.14$  \\
\hline
KMT-2018-BLG-0247 & 1.0 & 08 Jul 2018 & 17:38:14.41 & $-27$:09:01.48  & $+0.66$  & $+2.33$  \\
OGLE-2018-BLG-1219& 1.3 \\  
\hline
OGLE-2018-BLG-0298& 1.0 & 05 Mar 2018 & 17:37:08.28  & $-29$:42:32.80  & $-1.63$  & $+1.16$  \\
KMT-2018-BLG-1354 & 1.0 \\
\hline
KMT-2018-BLG-2602& 0.4 & Post Season & 17:49:35.29  & $-21$:58:34.32  & $+6.43$ & $+2.83$  \\ 
\hline
OGLE-2018-BLG-1119& 0.3 & 22 Jun 2018 & 18:00:07.02 & $-32$:22:31.0  & $-1.38$  & $-4.43$  \\
KMT-2018-BLG-1870& 0.4 \\
\hline
KMT-2018-BLG-0173 & 0.4 & 21 Jun 2018 & 17:50:11.75 & $-21$:35:40.56  & $+6.83$  & $+2.91$  \\
\hline
KMT-2018-BLG-1497 & 1.0 & Post Season & 17:44:20.19  & $-25$:58:25.00  & $+2.38$  & $+1.79$  \\
\hline
KMT-2018-BLG-1714 & 1.0 & Post Season & 17:50:27.27  & $-33$:22:33.82  & $-3.27$  & $-3.17$  \\
\hline
 \enddata
 \label{tab:names}
 \end{deluxetable}

 \begin{deluxetable}{lc}
 \tablecolumns{2} \tablewidth{0pc}
\tablecaption{\textsc{Light Curve Parameters for KMT-2018-BLG-0030 }}
 \tablehead{\colhead{Parameter} }
 \startdata
$\chi^2/$dof                  & $ 2397.43/ 2388$\\
$t_0 -8270$                   & $  1.459\pm  0.043$\\
$u_0\ (10^{-2})$              & $  90.05\pm   0.15$\\
$t_\e$ (days)                 & $  27.94\pm   0.11$\\
$s$                           & $ 1.580\pm 0.013$\\
$q\ (10^{-3})$                & $  2.74\pm  0.30$\\
$\langle\log q\rangle$        & $ -2.563\pm  0.048$\\
$\alpha$ (rad)                & $ 2.3521\pm 0.0065$\\
$\rho\ (10^{-3})$             & $< 112$\\
$I_{\rm S}$                   & $  16.82\pm   0.00$\\
 \enddata
 \label{tab:kb0030parms}
 \end{deluxetable}

 \begin{deluxetable}{lcc}
 \tablecolumns{3} \tablewidth{0pc}
\tablecaption{\textsc{Light Curve Parameters for KMT-2018-BLG-0087 }}
 \tablehead{\colhead{Parameter} &
 \colhead{Inner} & \colhead{Outer}}
 \startdata
$\chi^2/$dof                  & $ 3103.66/ 3087$& $ 3098.47/ 3087$\\
$t_0 -8280$                   & $ 1.7373\pm 0.0077$& $ 1.7290\pm 0.0079$\\
$u_0\ (10^{-2})$              & $ 51.4\pm  1.7$& $ 52.8\pm  1.8$\\
$t_\e$ (days)                 & $  4.607\pm  0.088$& $  4.536\pm  0.090$\\
$s$                           & $  0.638\pm  0.014$& $  0.898\pm  0.024$\\
$q\ (10^{-3})$                & $  2.73\pm  0.51$& $  2.17\pm  0.46$\\
$\langle\log q\rangle$        & $ -2.569\pm  0.083$& $ -2.668\pm  0.092$\\
$\alpha$ (rad)                & $  5.001\pm  0.012$& $  4.986\pm  0.012$\\
$\rho\ (10^{-3})$             & $<  96$& $< 110$\\
$I_{\rm S}$                   & $  16.86\pm   0.05$& $  16.82\pm   0.05$\\
 \enddata
 \label{tab:kb0087parms}
 \end{deluxetable}

 \begin{deluxetable}{lcc}
 \tablecolumns{3} \tablewidth{0pc}
\tablecaption{\textsc{Light Curve Parameters for KMT-2018-BLG-0247 }}
 \tablehead{\colhead{Parameter} &
 \colhead{Close} & \colhead{Wide}}
 \startdata
$\chi^2/$dof                  & $ 4172.76/ 4854$& $ 4171.10/ 4854$\\
$t_0 -8300$                   & $ 8.4239\pm 0.0065$& $ 8.4206\pm 0.0062$\\
$u_0\ (10^{-2})$              & $  6.86\pm  0.40$& $  6.51\pm  0.36$\\
$t_\e$ (days)                 & $ 10.56\pm  0.47$& $ 10.67\pm  0.47$\\
$s$                           & $ 0.9720\pm 0.0045$& $ 1.1182\pm 0.0048$\\
$q\ (10^{-3})$                & $  6.28\pm  0.46$& $  7.11\pm  0.56$\\
$\langle\log q\rangle$        & $ -2.203\pm  0.032$& $ -2.149\pm  0.034$\\
$\alpha$ (rad)                & $  1.146\pm  0.014$& $  1.200\pm  0.010$\\
$\rho\ (10^{-3})$             & $   1.92\pm   0.19$& $   2.52\pm   0.26$\\
$I_{\rm S}$                   & $  20.95\pm   0.06$& $  20.96\pm   0.06$\\
 \enddata
 \label{tab:kb0247parms}
 \end{deluxetable}

 \begin{deluxetable}{lccc}
 \tablecolumns{4} \tablewidth{0pc}
\tablecaption{\textsc{Light Curve Parameters for OGLE-2018-BLG-0298}}
 \tablehead{\colhead{Parameter} &
 \colhead{Close} & \colhead{Wide} &
 \colhead{1L2S}}
 \startdata
$\chi^2/$dof                  & $ 3116.68/ 3258$& $ 3119.27/ 3258$& $ 3150.36/ 3257$\\
$t_0 -8180$                   & $ 8.7417\pm 0.0052$& $ 8.7387\pm 0.0052$& $ 8.7078\pm 0.0048$\\
$t_{0,2} -8180$                &  & & $ 10.646\pm  0.014$\\
$u_0\ (10^{-2})$              & $  2.151\pm  0.077$& $  2.149\pm  0.077$& $  2.317\pm  0.139$\\
$u_{0,2}\ (10^{-2})$           &  & & $  0.147\pm  0.094$\\
$t_\e$ (days)                 & $  32.08\pm   0.98$& $  32.28\pm   0.99$& $  33.35\pm   1.11$\\
$s$                           & $  0.957\pm  0.018$& $  1.079\pm  0.016$& \\
$q\ (10^{-3})$                & $  0.199\pm  0.045$& $  0.137\pm  0.027$& \\
$\langle\log q\rangle$        & $ -3.705\pm  0.099$& $ -3.861\pm  0.081$& \\
$\alpha$ (rad)                & $  0.350\pm  0.006$& $  0.357\pm  0.005$& \\
$\rho\ (10^{-3})$             & $   3.58\pm   0.73$& $   3.98\pm   0.48$& $<25.50$\\
 $\rho_2\ (10^{-3})$            & & & $6.20\pm 0.59$\\
$q_F\ (10^{-3})$                   &  & & $ 13.1\pm 1.4$\\
$I_{\rm S}$                   & $  20.67\pm   0.04$& $  20.68\pm   0.04$& $  20.74\pm   0.04$\\
 \enddata
 \label{tab:ob0298parms}
 \end{deluxetable}

 \begin{deluxetable}{lccc}
 \tablecolumns{4} \tablewidth{0pc}
\tablecaption{\textsc{Light Curve Parameters for KMT-2018-BLG-2602 }}
 \tablehead{\colhead{Parameter} &
 \colhead{Inner} & \colhead{Outer} &
 \colhead{1L2S}}
 \startdata
$\chi^2/$dof                  & $  936.65/  920$& $  926.34/  920$& $  957.01/  919$\\
$t_0 -8260$                   & $ 9.95\pm 0.14$& $10.33\pm 0.15$& $11.24\pm 0.18$\\
$t_{0,2} -8260$                &  & & $-16.59\pm  0.29$\\
$u_0\ (10^{-2})$              & $ 56.5\pm 11.4$& $ 51.8\pm  9.3$& $ 46.9\pm  8.6$\\
$u_{0,2}\ (10^{-2})$           &  & & $  0.092\pm  0.513$\\
$t_\e$ (days)                 & $  94.4\pm  16.1$& $  98.7\pm  14.0$& $ 108.9\pm  13.7$\\
$s$                           & $  1.532\pm  0.084$& $  1.182\pm  0.065$& \\
$q\ (10^{-3})$                & $  1.62\pm  0.25$& $  1.65\pm  0.27$& \\
$\langle\log q\rangle$        & $ -2.794\pm  0.067$& $ -2.782\pm  0.071$& \\
$\alpha$ (rad)                & $  2.033\pm  0.017$& $  2.057\pm  0.015$& \\
$\rho\ (10^{-3})$             & $< 72$& $< 74$& \\
 $\rho_2\ (10^{-3})$            & & & $35.8\pm 6.1$\\
$q_F\ (10^{-3})$                   &  & & $ 3.44\pm 0.57$\\
$I_{\rm S}$                   & $  17.88\pm   0.33$& $  18.03\pm   0.28$& $  18.21\pm   0.26$\\
 \enddata
 \label{tab:kb2602parms}
 \end{deluxetable}

 \begin{deluxetable}{lccc}
 \tablecolumns{4} \tablewidth{0pc}
\tablecaption{\textsc{Light Curve Parameters for OGLE-2018-BLG-1119}}
 \tablehead{\colhead{Parameter} &
 \colhead{Inner} & \colhead{Outer} &
 \colhead{1L2S}}
 \startdata
$\chi^2/$dof                  & $ 1205.16/ 1267$& $ 1210.73/ 1267$& $ 1219.17/ 1266$\\
$t_0 -8310$                   & $ 5.98\pm 0.13$& $ 6.05\pm 0.13$& $ 6.28\pm 0.14$\\
$t_{0,2} -8310$                &  & & $  0.87\pm  0.16$\\
$u_0\ (10^{-2})$              & $ 43.5\pm  5.2$& $ 40.5\pm  4.6$& $ 43.2\pm  5.4$\\
$u_{0,2}\ (10^{-2})$           &  & & $  0.59\pm  0.78$\\
$t_\e$ (days)                 & $  39.3\pm   3.2$& $  41.2\pm   3.3$& $  40.3\pm   3.4$\\
$s$                           & $  1.426\pm  0.047$& $  1.081\pm  0.038$& \\
$q\ (10^{-3})$                & $  1.81\pm  0.46$& $  1.70\pm  0.43$& \\
$\langle\log q\rangle$        & $ -2.74\pm  0.11$& $ -2.78\pm  0.11$& \\
$\alpha$ (rad)                & $  1.857\pm  0.016$& $  1.869\pm  0.016$& \\
$\rho\ (10^{-3})$             & $< 71$& $< 61$& \\
 $\rho_2\ (10^{-3})$            & & & $34.8\pm 6.8$\\
$q_F\ (10^{-3})$                 &  & & $ 4.2\pm 1.0$\\
$I_{\rm S}$                   & $  19.53\pm   0.18$& $  19.64\pm   0.17$& $  19.58\pm   0.19$\\
 \enddata
 \label{tab:ob1119parms}
 \end{deluxetable}

 \begin{deluxetable}{lccc}
 \tablecolumns{4} \tablewidth{0pc}
\tablecaption{\textsc{Light Curve Parameters for KMT-2018-BLG-0173 }}
 \tablehead{\colhead{Parameter} &
 \colhead{Close} & \colhead{Wide} &
 \colhead{1L2S}}
 \startdata
$\chi^2/$dof                  & $  927.58/  921$& $  938.79/  921$& $  938.44/  921$\\
$t_0 -8340$                   & $ 8.74\pm 0.11$& $ 8.65\pm 0.11$& $ 8.71\pm 0.11$\\
$t_{0,2} -8340$                &  & & $-83.74\pm  0.15$\\
$u_0\ (10^{-2})$              & $ 62.0\pm  4.1$& $ 79.7\pm  5.7$& $ 83.0\pm  1.2$\\
$u_{0,2}\ (10^{-2})$           &  & & $  1.87\pm  0.89$\\
$t_\e$ (days)                 & $  62.51\pm   2.90$& $  52.17\pm   2.59$& $  50.48\pm   0.37$\\
$s$                           & $  0.478\pm  0.015$& $  2.314\pm  0.082$& \\
$q\ (10^{-3})$                & $  1.05\pm  0.17$& $  0.97\pm  0.27$& \\
$\langle\log q\rangle$        & $ -2.981\pm  0.069$& $ -3.021\pm  0.125$& \\
$\alpha$ (rad)                & $ 5.7990\pm 0.0114$& $ 2.7752\pm 0.0087$& \\
$\rho\ (10^{-3})$             & $< 44$& $ 125\pm 19$& \\
 $\rho_2\ (10^{-3})$          &  & & $42.5\pm 7.3$\\
$q_F\ (10^{-3})$              &  & & $ 1.04\pm 0.15$\\
$I_{\rm S}$                   & $  17.46\pm   0.12$& $  16.98\pm   0.14$& $  16.90\pm   0.00$\\
 \enddata
 \label{tab:kb0173parms}
 \end{deluxetable}

 \begin{deluxetable}{lcccc}
 \tablecolumns{5} \tablewidth{0pc}
\tablecaption{\textsc{Light Curve Parameters for KMT-2018-BLG-1497 }}
 \tablehead{\colhead{Parameter} &
 \colhead{Inner} & \colhead{Outer}&\colhead{Off-axis Cusp}&
 \colhead{1L2S}}
 \startdata
$\chi^2/$dof                  & $ 2466.85/ 2457$& $ 2463.63/ 2457$& $ 2467.78/ 2457$& $ 2465.74/ 2456$\\
$t_0 -8220$                   & $ 9.22\pm 0.11$& $ 9.14\pm 0.11$& $ 8.07\pm 0.12$& $ 8.98\pm 0.12$\\
$t_{0,2} -8220$                &  &  & & $ 13.854\pm  0.026$\\
$u_0\ (10^{-2})$              & $ 22.1\pm  2.3$& $ 21.0\pm  1.8$& $ 18.4\pm  1.3$& $ 19.2\pm  2.4$\\
$u_{0,2}\ (10^{-2})$           &  &  & & $  0.23\pm  0.14$\\
$t_\e$ (days)                 & $  30.6\pm   2.3$& $  31.9\pm   2.0$& $  34.0\pm   1.9$& $  34.2\pm   2.9$\\
$s$                           & $  1.227\pm  0.019$& $  1.128\pm  0.012$& $  0.928\pm  0.005$& \\
$q\ (10^{-3})$                & $  0.75\pm  0.24$& $  0.21\pm  0.07$& $ 17.38\pm  2.48$& \\
$\langle\log q\rangle$        & $ -3.127\pm  0.136$& $ -3.684\pm  0.139$& $ -1.763\pm  0.063$& \\
$\alpha$ (rad)                & $ 0.971\pm 0.020$& $ 0.955\pm 0.019$& $ 2.971\pm 0.045$& \\
$\rho\ (10^{-3})$             & $<  8.6$& $7.8\pm 1.7$& $<  7.8$& \\
 $\rho_2\ (10^{-3})$           &  &  & & $<6.7$       \\
$q_F\ (10^{-3})$               &  &  & & $ 4.2\pm 1.6$\\
$I_{\rm S}$                   & $  20.79\pm   0.13$& $  20.87\pm   0.11$& $  21.03\pm   0.09$& $  21.00\pm   0.15$\\
 \enddata
 \label{tab:kb1497parms}
 \end{deluxetable}

 \begin{deluxetable}{lcccc}
 \tablecolumns{5} \tablewidth{0pc}
\tablecaption{\textsc{Light Curve Parameters for KMT-2018-BLG-1714 }}
 \tablehead{\colhead{Parameter} &
 \colhead{Inner} & \colhead{Outer}&\colhead{Off-axis Cusp}&
 \colhead{1L2S}}
 \startdata
$\chi^2/$dof                  & $ 2033.08/ 2028$& $ 2032.94/ 2028$& $ 2040.35/ 2028$& $ 2033.80/ 2027$\\
$t_0 -8310$                   & $ 8.168\pm 0.012$& $ 8.170\pm 0.012$& $ 8.280\pm 0.015$& $ 8.179\pm 0.014$\\
$t_{0,2} -8310$                &  &  & & $  8.1047\pm  0.0025$\\
$u_0\ (10^{-2})$              & $ 15.7\pm  2.0$& $ 16.2\pm  2.2$& $ 14.4\pm  1.3$& $ 18.9\pm  3.1$\\
$u_{0,2}\ (10^{-2})$           &  &  & & $  0.056\pm  0.184$\\
$t_\e$ (days)                 & $   3.21\pm   0.29$& $   3.17\pm   0.30$& $   3.52\pm   0.22$& $   3.12\pm   0.28$\\
$s$                           & $  1.293\pm  0.038$& $  0.921\pm  0.028$& $  0.974\pm  0.004$& \\
$q\ (10^{-3})$                & $  3.97\pm  1.67$& $  3.68\pm  1.74$& $ 12.83\pm  2.43$& \\
$\langle\log q\rangle$        & $ -2.40\pm  0.14$& $ -2.43\pm  0.15$& $ -1.89\pm  0.081$& \\
$\alpha$ (rad)                & $ 1.693\pm 0.024$& $ 1.694\pm 0.024$& $ 0.030\pm 0.042$& \\
$\rho\ (10^{-3})$             & $< 15.6$& $< 16.8$& $< 10.3$& \\
 $\rho_2\ (10^{-3})$            & & & &$<12.4$\\
$q_F\ (10^{-3})$               &  &  & & $13.5\pm 2.6$\\
$I_{\rm S}$                   & $  20.56\pm   0.17$& $  20.54\pm   0.17$& $  20.72\pm   0.12$& $  20.50\pm   0.19$\\
 \enddata
 \label{tab:kb1714parms}
 \end{deluxetable}

 \begin{deluxetable}{lrrrrrrrr}
 \tablecolumns{9} \rotate \tablewidth{0pc}
 \tablecaption{\textsc {CMD Parameters}}
 \tablehead{\colhead{Name} & 
\colhead{$(V-I)_{\rm S}$} &
\colhead{$(V-I)_{\rm cl}$} &
\colhead{$(V-I)_{\rm S,0}$} &
\colhead{$I_{\rm S}$} &
\colhead{$I_{\rm cl}$} &
\colhead{$I_{\rm cl,0}$} &
\colhead{$I_{\rm S,0}$} &
\colhead{$\theta_*\ (\muas)$} }
 \startdata
KMT-2018-BLG-0030&  3.57$\pm$0.03 & 3.49$\pm$0.03 & 1.14$\pm$0.04 & 16.88$\pm$0.01 & 17.40$\pm$0.03 & 14.45 & 13.93$\pm$0.03 & 8.478$\pm$0.526 \\
KMT-2018-BLG-0087&  3.98$\pm$0.05 & 3.67$\pm$0.03 & 1.37$\pm$0.06 & 16.97$\pm$0.02 & 17.45$\pm$0.03 & 14.44 & 13.96$\pm$0.05 & 9.529$\pm$0.632 \\
KMT-2018-BLG-0247&  3.11$\pm$0.06 & 3.65$\pm$0.02 & 0.52$\pm$0.07 & 21.03$\pm$0.03 & 17.43$\pm$0.03 & 14.41 & 18.01$\pm$0.04 & 0.641$\pm$0.055 \\
 OGLE-2018-BLG-0298& 4.36$\pm$0.06 & 4.65$\pm$0.03 & 0.77$\pm$0.07 & 20.74$\pm$0.03 & 18.55$\pm$0.06 & 14.53 & 16.72$\pm$0.07 & 1.513$\pm$0.142 \\
KMT-2018-BLG-2602&  2.93$\pm$0.07 & 2.93$\pm$0.03 & 1.06$\pm$0.08 & 18.22$\pm$0.03 & 16.60$\pm$0.05 & 14.26 & 15.88$\pm$0.06 & 3.207$\pm$0.310 \\
OGLE-2018-BLG-1119& 1.64$\pm$0.11 & 2.00$\pm$0.03 & 0.70$\pm$0.12 & 19.50$\pm$0.03 & 15.73$\pm$0.05 & 14.52 & 18.29$\pm$0.06 & 0.681$\pm$0.095 \\
KMT-2018-BLG-0173&  3.01$\pm$0.06 & 3.05$\pm$0.03 & 1.02$\pm$0.07 & 17.67$\pm$0.02 & 16.96$\pm$0.08 & 14.25 & 14.96$\pm$0.08 & 4.371$\pm$0.371 \\
KMT-2018-BLG-1497&  3.13$\pm$0.10 & 3.26$\pm$0.03 & 0.93$\pm$0.11 & 20.96$\pm$0.03 & 17.45$\pm$0.04 & 14.36 & 17.87$\pm$0.05 & 1.110$\pm$0.166 \\
KMT-2018-BLG-1714&  2.60$\pm$0.11 & 2.41$\pm$0.03 & 1.25$\pm$0.12 & 20.08$\pm$0.03 & 16.25$\pm$0.05 & 14.59 & 18.42$\pm$0.06 & 1.226$\pm$0.114 \\
\hline
 \enddata
 \tablecomments{$(V-I)_{\rm cl,0}=1.06$}
 \label{tab:cmd}
 \end{deluxetable}


\begin{deluxetable}{lccccccc}
\tablecolumns{8} 
\tablewidth{0pc}\tablecaption{\textsc{Physical properties}} 
\tablehead{\colhead{Event} & \multicolumn{4}{c}{Physical Parameters} & \colhead{} &
\multicolumn{2}{c}{Relative Weights}\\
\cline{1-1} \cline{7-8} 
\colhead{Models}
& \colhead{$M_{\rm host}$ $[M_\sun]$}  
& \colhead{$M_{\rm planet}$ $[M_{\rm Jup}]$}  
& \colhead{$D_{\rm L}$ [kpc]}
& \colhead{$a_\bot$ [au]} &
& \colhead{Gal.Mod.} & \colhead{$\chi^2$}} \startdata

KB180030 & $0.51^{+0.43}_{-0.31}$ & $1.45^{+1.23}_{-0.88}$ & $6.48^{+1.28}_{-1.96}$ & $4.39^{+2.18}_{-2.40}$ & & 1.00 & 1.00\\
 \cline{1-8}
KB180087 \\
Inner        & $0.11^{+0.15}_{-0.06}$ & $0.32^{+0.43}_{-0.16}$ & $6.90^{+1.04}_{-1.16}$ & $0.66^{+0.20}_{-0.21}$ & & 0.62 & 0.07\\
Outer        & $0.10^{+0.14}_{-0.05}$ & $0.23^{+0.32}_{-0.12}$ & $7.02^{+1.03}_{-1.15}$ & $0.87^{+0.24}_{-0.25}$ & & 1.00 & 1.00 \\
{\bf Adopted}& $0.10^{+0.14}_{-0.05}$ & $0.23^{+0.32}_{-0.12}$ & $7.02^{+1.03}_{-1.15}$ & $0.87^{+0.24}_{-0.25}$ \\
 \cline{1-8}
KB180247 \\
Close        & $0.35^{+0.31}_{-0.18}$ & $2.33^{+2.07}_{-1.18}$ & $6.47^{+0.99}_{-1.33}$ & $2.11^{+0.43}_{-0.52}$ & & 0.65 & 0.44 \\
Wide         & $0.27^{+0.28}_{-0.14}$ & $2.04^{+2.09}_{-1.04}$ & $6.84^{+0.99}_{-1.24}$ & $2.56^{+0.51}_{-0.58}$ & & 1.00 & 1.00 \\
{\bf Adopted}& $0.29^{+0.28}_{-0.14}$ & $2.11^{+2.09}_{-1.04}$ & $6.76^{+0.99}_{-1.24}$ & $2.46^{+0.51}_{-0.58}$ \\
 \cline{1-8}
OB180298 \\
Close        & $0.70^{+0.34}_{-0.30}$ & $0.15^{+0.07}_{-0.06}$ & $6.49^{+0.95}_{-1.23}$ & $2.84^{+0.73}_{-0.80}$ & & 1.00 & 1.00 \\
Wide         & $0.63^{+0.34}_{-0.27}$ & $0.091^{+0.049}_{-0.038}$ & $6.79^{+0.89}_{-1.07}$ & $2.98^{+0.54}_{-0.59}$& &0.91 & 0.27 \\
{\bf Adopted}& $0.69^{+0.34}_{-0.30}$ & $0.14^{+0.07}_{-0.06}$ & $6.54^{+0.95}_{-1.23}$ & $2.86^{+0.73}_{-0.80}$ \\
 \cline{1-8}
KB182602 & $0.66^{+0.42}_{-0.36}$ & $1.15^{+0.73}_{-0.63}$ & $4.31^{+1.97}_{-1.84}$ & $3.81^{+2.96}_{-2.32}$ & & 1.00 & 1.00 \\
 \cline{1-8}
OB181119 \\
Inner        & $0.48^{+0.35}_{-0.28}$ & $0.91^{+0.66}_{-0.52}$ & $5.76^{+1.43}_{-2.48}$ & $4.11^{+2.14}_{-2.58}$ & & 1.00 & 1.00 \\
Outer        & $0.48^{+0.35}_{-0.28}$ & $0.86^{+0.62}_{-0.49}$ & $5.70^{+1.47}_{-2.48}$ & $3.13^{+1.65}_{-1.98}$ & & 0.89 & 0.06 \\
{\bf Adopted}& $0.48^{+0.35}_{-0.28}$ & $0.91^{+0.66}_{-0.52}$ & $5.76^{+1.43}_{-2.48}$ & $4.06^{+2.14}_{-2.58}$ \\
 \cline{1-8}
 \enddata
 \label{tab:physall}
\end{deluxetable}

\begin{deluxetable}{llccl}
\tablecolumns{5} \tablewidth{0pc}
\tablecaption{\textsc{AnomalyFinder Planets in KMT Sub-prime Fields for 2018}}
\tablehead{\colhead{Event Name} &
\colhead{KMT Name} &
\colhead{$\log q$} &
\colhead{$s$} &
\colhead{Reference} }
\startdata
KB180029       & KB180029 & $-4.74$ & 1.00 & \citet{kb180029} \\
OB180298$^{AA}$ & KB181354 & $-3.71$ & 0.96 & This Work \\
KB181996$^{DD}$ & KB181996 & $-2.82$ & 1.46 & \citet{kb181976} \\
KB182602       & KB182602 & $-2.78$ & 1.18 & This Work \\
OB181428       & KB180423 & $-2.76$ & 1.42 & \citet{ob181428} \\
OB181119$^{AA}$ & KB181870 & $-2.74$ & 0.64 & This Work \\
KB180087$^{AA}$ & KB180087 & $-2.67$ & 2.17 & This Work \\
OB180799       & KB181741 & $-2.60$ & 1.13 & \citet{ob180799} \\
KB180030       & KB180030 & $-2.56$ & 1.58 & This Work \\
KB181976$^{AA}$ & KB181976 & $-2.50$ & 1.23 & \citet{kb181976} \\
KB181292       & KB181292 & $-2.45$ & 1.36 & \citet{kb181292} \\
KB181990$^{AA,BB}$&KB181990 & $-2.45$ & 0.96 & \citet{kb181990} \\
OB180740$^{AA}$ & KB181822 & $-2.34$ & 1.26 & \citet{ob180740} \\
KB180247$^{AA}$ & KB180247 & $-2.15$ & 1.12 & This Work \\
\hline
\hline
KB181988$^{AA,GG}$  & KB181988 & $-4.45$ & 1.04 & \citet{kb181988} \\
KB181497$^{AA,CC,GG}$& KB181497 & $-3.68$ & 1.13 & This Work \\
KB180173$^{AA,CC}$  & KB180173 & $-2.98$ & 0.48 & This Work \\
KB181743$^{AA,GG}$  & KB181743 & $-2.92$ & 1.05 &\citet{kb181743}\\
KB181714$^{AA,CC,GG}$& KB181714 & $-2.43$ & 0.92 & This Work \\
\enddata
\tablecomments{Event names are abbreviations for, e.g.,
KMT-2018-BLG-0029 and OGLE-2018-BLG-0799.
AA: $s$ degeneracy. 
BB: Factor 1.6 $q$ degeneracy.
CC: 1L2S/2L1S degeneracy.
DD: Not selected in AnomalyFinder review.
GG: large $q$ degeneracy.
}
\label{tab:all2018events}
\end{deluxetable}

\clearpage

\begin{thebibliography}{99}

\bibitem[Alard \& Lupton(1998)]{alard98} Alard, C. \& Lupton, R.H.,1998, \apj, 503, 325

\bibitem[Albrow(2017)]{pydia}Albrow, M.D. Michaeldalbrow/Pydia: InitialRelease On Github., vv1.0.0, Zenodo

\bibitem[Albrow et al.(2009)]{albrow09}Albrow, M.\ D., Horne, K., Bramich, D.\ M., et al.\ 2009, \mnras, 397, 2099






\bibitem[Bensby et al.(2013)]{bensby13} Bensby, T. Yee, J.C., Feltzing, S.\ et al.\ 2013, \aap, 549, A147

\bibitem[Bessell \& Brett(1988)]{bb88} Bessell, M.S., \& Brett, J.M.\ 1988, \pasp, 100, 1134







\bibitem[Gaia Collaboration et al.(2016)]{gaia16}Gaia Collaboration, Prusti, T., de Bruijne, J.H.J., et al. 2016, \aap, 595, A1

\bibitem[Gaia Collaboration et al.(2018)]{gaia18}Gaia Collaboration, Brown, A. G. A., Vallenari, A., et al.\ 2018, \aap, 616, 1


\bibitem[Gaudi(1998)]{gaudi98} Gaudi, B.S.\ 1998, \apj, 506, 533

\bibitem[Gaudi \& Gould(1997)]{gaudi97} Gaudi, B.S. \& Gould, A.\ 1997, \apj, 486, 85


\bibitem[Gonzalez et al.(2012)]{gonzalez12} Gonzalez, O.~A., Rejkuba, M., Zoccali, M., et al.\ 2012, \aap, 543, A13







\bibitem[Gould et al.(2021)]{gould21} Gould, A., Zang, W., Mao, S., \& Dong, S., 2021, RAA, 21, 133 

\bibitem[Gould et al.(2022)]{2018prime} Gould, A., Han, C., Zang, W.., 2022, \aap, in press, arXiv: arXiv:2204.04354

\bibitem[Gould et al.(2020)]{kb180029} Gould, A., Ryu, Y.-H., Calchi Novati, S., et al.\ 2020, JKAS, 53, 9

\bibitem[Griest \& Safizadeh(1998)]{griest98} Griest, K.\ \& Safizadeh, N.\ 1998, \apj, 500, 37



\bibitem[Han et al.(2016)]{ob150479}Han, C., Udalski, A., Gould, A., et al. 2016, \apj, 828, 53


\bibitem[Han et al.(2019)]{ob180740} Han, C., Yee, J.C., Udalski, A., et al.\ 2019, \aj, 158, 102






\bibitem[Han et al.(2021a)]{kb181976}Han, C., Udalski, A., Kim, D., et al. 2021a, \aap, 650A, 89

\bibitem[Han et al.(2021b)]{kb181743}Han, C., Albrow, M.D., Chung, S.-J., et al. 2021b, \aap, 652A, 145

\bibitem[Han et al.(2022c)]{kb181988}Han, C., Gould, A., Albrow, M.D., et al. 2021c, \aap, 658A, 62

\bibitem[Henderson et al.(2014)]{henderson14} Henderson, C.B., Gaudi, B.S., Han, C., et al. 2014, \apj, 794, 52







\bibitem[Hwang et al.(2022)]{kb190253} Hwang, K.-H., Zang, W., Gould, A., et al.., 2022, \aj, 163, 43 





\bibitem[Kervella et al.(2004a)]{kervella04a} Kervella, P., Bersier, D., Mourard, D., et al.\ 2004a, \aap, 428, 587

\bibitem[Kervella et al.(2004b)]{kervella04b} Kervella, P., Th{\'e}venin, F., Di Folco, E., \& S{\'e}gransan, D.\ 2004b, \aap, 426, 297

\bibitem[Kim et al.(2016)]{kmtnet} Kim, S.-L., Lee, C.-U., Park, B.-G., et al.  2016, JKAS, 49, 37 

\bibitem[Kim et al.(2018a)]{eventfinder} Kim, D.-J., Kim,  H.-W., Hwang, K.-H., et al., 2018a, \aj, 155, 76

\bibitem[Kim et al.(2018b)]{alertfinder} Kim, H.-W., Hwang, K.-H., Shvartzvald, Y., et al. 2018b, arXiv:1806.07545


\bibitem[Kim et al.(2021a)]{ob181428}Kim, Y.-H., Chung, S.-J., Udalski, A., et al. 2021a, \mnras, 503, 2706

\bibitem[Kim et al.(2021b)]{kb190371}Kim, Y.H.., Chung, S.-J., Yee, J.-C., et al. 2021b, \aj, 162, 17


\bibitem[Kubiak \& Szyma\'nski(1997)]{oiicat2} Kubiak, M. \& Szymański, M.K.\ 1997, Acta Astron., 47, 319. 

\bibitem[Minniti et al.(2010)]{vvv-survey1}Minniti, D., Lucas, P. W., Emerson, J. P., et al. 2010, New Astron., 15, 433

\bibitem[Minniti et al.(2017)]{vvvcat}Minniti, D., Lucas, P., VVV Team, 2017, yCAT 2348, 0

\bibitem[Nataf et al.(2013)]{nataf13} Nataf, D.M., Gould, A., Fouqu\'e, P. et al. 2013, \apj, 769, 88

\bibitem[Paczy\'nski(1986)]{pac86} Paczy\'nski, B.\ 1986, \apj, 304, 1

\bibitem[Park et al.(2004)]{mb03037}Park, B.-G., DePoy, D.L.., Gaudi, B.S.,  et al.\ 2004, \apj, 609, 166

\bibitem[Ryu et al.(2019)]{kb181990} Ryu, Y.-H., Hwang, K.-H., Gould, A. et al. 2019, \aj, 158, 151

\bibitem[Ryu et al.(2020)]{kb181292} Ryu, Y.-H., Navarro, M.G., Gould, A. et al. 2020, \aj, 159, 58



\bibitem[Ryu et al.(2021)]{kb162605} Ryu, Y.-H., Hwang, K.-H., Gould, A. et al. 2021, \aj, 162, 96

\bibitem[Ryu et al.(2022)]{kb211391} Ryu, Y.-H., Jung, Y.K., Yang, H., et al. 2022, \aj, submitted, arXiv:2202.03022






\bibitem[Szyma\'nski(2005)]{oiicat1}Szyma\'nski, M.K.\ 2005, Acta Astron., 55, 43

\bibitem[Szyma\'nski et al.(2011)]{oiiicat}Szyma\'nski, M.K., Udalski, A., Soszy\'nski, I., et al. 2011, Acta Astron., 61, 83

\bibitem[Tomaney \& Crotts(1996)]{tomaney96} Tomaney, A.B. \& Crotts, A.P.S. 1996, \au, 112, 2872

\bibitem[Udalski et al.(2002)]{oiicat3} Udalski, A. Szyma\'nski, M., Kubiak, M., et al., 2002, Acta Astron., 52, 217

\bibitem[Udalski(2003)]{ews2} Udalski, A. 2003, Acta Astron., 53, 291

\bibitem[Udalski et al.(1994)]{ews1} Udalski, A.,Szymanski, M., Kaluzny, J., et al.\ 1994, Acta Astron., 44, 227

\bibitem[Wo\'zniak(2000)]{wozniak2000} Wo\'zniak, P.~R. 2000, Acta Astron., 50, 421

\bibitem[Wang et al.(2022)]{ob180383} Wang, H., Zang, W., Zhu, W, et al.\ 2022, \mnras, 510, 1778


\bibitem[Yee et al.(2015)]{yee15} Yee, J.C., Gould, A., Beichman, C., 2015, \apj, 810, 155

\bibitem[Yee et al.(2021)]{ob190960} Yee, J.C., Zang, W., Udalski, A.\ et al. 2021, \aj, 162, 180 

\bibitem[Yoo et al.(2004)]{ob03262} Yoo, J., DePoy, D.L., Gal-Yam, A.\ et al.\ 2004, \apj, 603, 139


\bibitem[Zang et al.(2019)]{ob180799} Zang, W., Shvartzvald, Y., Udalski, A., et al. 2019, arXiv:2010.08732

\bibitem[Zang et al.(2021b)]{ob191053} Zang, W., Hwang, K.-H., Udalski, A., et al. 2021b, \aj, 162, 163 

\bibitem[Zang et al.(2022)]{af2} Zang, W., Yang, H., Han, c., et al.\ 2022, arXiv:2204.02017

\bibitem[Zhang \& Gaudi(2022)]{zhang22} Zhang, K. \& Gaudi, B.S.\ 2022, arXiv:2205.05085

\bibitem[Zhu et al.(2014)]{zhu14} Zhu, W., Penny, M., Mao, S., Gould, A., \& Gendron, R.\ 2014, \apj, 788, 73

\end{thebibliography}
\end{document}